\documentclass[lettersize,journal]{IEEEtran}
\usepackage{amsmath,amsfonts}
\usepackage{algorithmic}
\usepackage{algorithm}
\usepackage{array}
\usepackage{textcomp}
\usepackage{stfloats}
\usepackage{url}
\usepackage{verbatim}
\usepackage{cite}
\usepackage{sharina}
\usepackage{setspace}
\usepackage{amsthm}
\usepackage{amssymb,amsmath,latexsym}
\usepackage{graphicx}
\usepackage{color,cite,times}
\usepackage{epstopdf}
\usepackage{multirow}
\usepackage{array}
\usepackage[caption=false,font=footnotesize]{subfig}
\usepackage{booktabs}
\usepackage{hhline}

\newtheorem{proposition}{\textbf{Proposition}}
\newtheorem{remark}{\textbf{Remark}}
\allowdisplaybreaks[4]

\begin{document}
	
	\title{
	Integrated Sensing and Communication with UAV Swarms via Decentralized Consensus ADMM}
	\author{	
		Zhiyuan~Zhai, Wei Ni,~\IEEEmembership{Fellow, IEEE},
	 Xin~Wang,~\IEEEmembership{Fellow, IEEE}, Dusit Niyato,~\IEEEmembership{Fellow, IEEE}, \\
     and Ekram Hossain,~\IEEEmembership{Fellow, IEEE}
}

	\maketitle


\begin{abstract}
    UAV swarms can form virtual antenna arrays to exploit additional spatial degrees of freedom and enhance integrated sensing and communication~(ISAC). The optimization of UAV positions is challenging due to the distributed nature of swarms and the lack of a global view at individual UAVs.     
    This paper presents a new decentralized optimization framework that allows UAVs to decide their locations in parallel and reach consensus on a globally optimal swarm geometry for ISAC.
	Specifically, we derive the achievable uplink rate and Cramér-Rao Bound (CRB) as tractable metrics for communication and sensing, respectively. 
	The UAV positions are optimized to balance maximizing the communication rate and minimizing the CRB.
	To solve this non-convex problem with coupled variables, we develop a decentralized consensus alternating direction method of multipliers (ADMM) algorithm, which enables the UAVs to iteratively align their local updates and reach consensus.
	The algorithm decomposes the global objective into local projection updates, proxy-assisted consensus coordination, and lightweight dual updates, ensuring scalability and consistency throughout the swarm. 
	Simulations demonstrate that the proposed consensus ADMM algorithm converges rapidly with strong scalability, and that the UAV swarm significantly outperforms fixed-array baselines in both communication and sensing performance.
\end{abstract}


\section{Introduction}

\subsection{Motivation and Challenges}

Integrated sensing and communication (ISAC)~\cite{lu2024integrated,liu2022integrated,11025997} has emerged as a key enabler for the sixth-generation communication systems, aiming to integrate high-capacity transmission and high-accuracy sensing within a unified platform. A fundamental requirement is to achieve both reliable communication and precise sensing, which are strongly influenced by the spatial configuration of the antenna array~\cite{lee2012spatial}. Traditional fixed arrays, such as uniform linear and planar arrays, offer limited spatial flexibility because their geometries are rigid and cannot adapt to time-varying environments or user locations. This rigidity restricts the achievable trade-off between communication throughput and sensing accuracy,  motivating research on {spatially reconfigurable or mobile arrays} that can dynamically adjust their geometry to optimize both functionalities.

A promising option is employing an unmanned aerial vehicle (UAV) \cite{10246260,9730072,10818523} swarm to form a virtual antenna array for ISAC~\cite{mozaffari2019tutorial,zeng2019accessing,liu2022isac,wei2023isacsignals}. 
In a UAV swarm, multiple UAVs cooperatively sense, communicate, and adapt their spatial configurations in three-dimensional (3D) space. 
Thanks to their mobility and controllability, UAVs can dynamically adjust their positions to construct a reconfigurable virtual array \cite{9891794,10100674}, enabling spatial diversity gains similar to, or even beyond, those of conventional large-scale antenna arrays~\cite{mozaffari2019tutorial}.  
In the context of ISAC, this mobility introduces additional degrees of freedom for reconfiguring the array geometry, hence enhancing channel spatial diversity for multiplexing, improving angular resolution for localization, and providing flexible trade-offs between communication and sensing~\cite{liu2022isac,wei2023isacsignals}.  
Moreover, the distributed nature of a UAV swarm allows for scalable, resilient, and adaptive operation without a centralized infrastructure, making it a promising architecture for next-generation intelligent networks.


Despite these advantages, deploying a UAV swarm to perform ISAC tasks is challenging. 
The communication rate and sensing accuracy are strongly coupled with the geometric configuration of the UAV swarm, which must be jointly optimized across all UAVs. 
However, each UAV only observes local information (e.g., its own channel or position), while the ISAC performance depends on the {global} spatial configuration of the entire swarm. 
This discrepancy creates a fundamental challenge: UAVs must coordinate to reach a common understanding of the optimal swarm geometry based on limited local views. 
Without effective consensus, inconsistent updates among UAVs can degrade the array coherence, leading to mismatched communication and sensing performance. 
Moreover, the resulting position optimization problem is highly non-convex and globally coupled due to the nonlinear dependence of both communication and sensing metrics on UAV positions, making centralized approaches impractical and distributed consensus nontrivial. 
These challenges call for new distributed frameworks that can achieve consensus while optimizing UAVs' positions, ensuring scalability and global consistency in UAV swarm–enabled ISAC systems.

\subsection{Related Work}
{
Multiple UAVs have been employed to cooperatively form a {virtual antenna array}  to enhance wireless communication performance. By adjusting their spatial positions, UAVs can emulate a large-scale distributed antenna system, providing additional spatial degrees of freedom and improving link capacity. For instance, Hanna \emph{et al.}~\cite{hanna2021uav} optimized UAV swarm geometry to maximize MIMO channel capacity in a backhaul scenario. Similarly, Shi \emph{et al.}~\cite{9771817} investigated air-to-air communications based on UAV-enabled virtual antenna arrays using a multi-objective optimization approach, while Li \emph{et al.}~\cite{10195219} optimized UAV-swarm-assisted IoT systems employing virtual antenna arrays to jointly enhance throughput and energy efficiency. These studies confirm the feasibility of using UAV swarms as reconfigurable virtual antenna arrays to strengthen communication links, yet they mainly focus on communication-only objectives without considering sensing functionality or distributed implementation.}

Recent years have witnessed growing interest in employing UAVs for ISAC. Jing~\emph{et~al.}~\cite{10529184} optimize the UAV trajectory to balance user rate and target localization accuracy through joint rate--CRB optimization, while Wu~\emph{et~al.}~\cite{10168298} design real-time UAV trajectories for secure ISAC under eavesdropping threats. Pang~\emph{et~al.}~\cite{10659350} propose dynamic beamforming for UAV-enabled vehicular ISAC, Khalili~\emph{et~al.}~\cite{10680299} consider joint hovering, resource allocation, and trajectory design with limited backhaul, Liu~\emph{et~al.}~\cite{10566041} address secure rate maximization in UAV-assisted ISAC, and Wu~\emph{et~al.}~\cite{10453349} investigate RIS-assisted UAV-enabled ISAC via joint trajectory and resource optimization.  
Although these studies demonstrate the potential of UAV-assisted ISAC, single-UAV platforms inherently suffer from limited spatial diversity and coverage since only one movable sensing and communication node is available, leading to restricted geometry adaptability and a tightly coupled rate-sensing trade-off.

To overcome the above limitations, recent works explore multi-UAV or UAV swarm-based ISAC systems. Zhang~\emph{et~al.}~\cite{10602493} jointly optimize trajectories, user association, and beamforming for multi-UAV-assisted ISAC, while Xie~\emph{et~al.}~\cite{10623895} develop a distributed UAV swarm ISAC framework using multi-agent reinforcement learning (MARL) for device-free sensing.  
These existing multi-UAV studies typically treat UAVs as independent nodes rather than a coordinated virtual antenna array, overlooking the global spatial gain of array cooperation. Moreover, their optimization frameworks rely on centralized global information, limiting their scalability for large-scale UAV swarms.  
These challenges motivate our exploration of the distributed optimization of a UAV swarm forming a virtual antenna array to jointly enhance communication and sensing performance. Table~\ref{tab:related_work_compact} compares the related works and this work.

\begin{table*}[t]
\centering
\caption{Comparison of representative UAV-assisted ISAC studies.}
\footnotesize
\renewcommand{\arraystretch}{1.05}
\setlength{\tabcolsep}{2pt}
\resizebox{\textwidth}{!}{%
\begin{tabular}{|l|p{9.5cm}|c|c|c|}
\hline
\textbf{Category / Reference} & \textbf{Main Contributions} & \textbf{UAV-ISAC} & \textbf{Multi-UAV} & \textbf{Distributed Opt.} \\
\hline
\raisebox{0pt}[0pt][0pt]{\textbf{Single-UAV ISAC} \cite{10529184,10168298,10659350,10680299,10566041,10453349}} 
& Studies on single-UAV ISAC with trajectory/beamforming/resource optimization; feasible but limited by single-node diversity and coverage. 
& \checkmark &  &  \\
\hline
\raisebox{0pt}[0pt][0pt]{\textbf{Multi-UAV ISAC} \cite{10602493,10623895}} 
& Multi-UAV (swarm) ISAC via cooperative control or MARL; improved joint communication–sensing, but mostly centralized and without array-level coordination. 
& \checkmark & \checkmark &  \\
\hline
\textbf{This paper} 
& UAV swarm forms a virtual antenna array and performs distributed geometry optimization for joint communication–sensing enhancement. 
& \checkmark & \checkmark & \checkmark \\
\hline
\end{tabular}}
\label{tab:related_work_compact}
\end{table*}

\subsection{Contributions}

This paper introduces a novel UAV swarm-enabled ISAC framework, where a group of UAVs forms a virtual antenna array that can be flexibly deployed in 3D space to jointly support communication and sensing. We analytically characterize the system performance by deriving the achievable uplink rate as the communication metric and the Cramér–Rao Bound (CRB) as a tractable measure of localization accuracy. Accordingly, we formulate a joint optimization problem that balances maximizing the communication rate and minimizing the CRB under practical spatial deployment constraints. To solve the resulting non-convex and tightly coupled problem, we develop a consensus-based alternating direction method of multipliers (ADMM) distributed optimization framework. This framework decomposes the global problem into local subproblems solved in closed form via gradient descent, thus ensuring scalability and convergence guarantees.

The key contributions of this paper include:

\begin{itemize}
	\item 
    We propose the concept of UAV swarm-enabled ISAC, where multiple UAVs cooperatively form a virtual antenna array. This extends the flexibility of fixed arrays and mechanically constrained movable antennas, enabling rich geometric configurations for ISAC.  

    \item 
    We analytically derive the achievable uplink rate and CRB of the UAV swarm-enabled ISAC system as the geometry-aware communication performance metric and a measure of localization accuracy, respectively.
    
	\item 
    We establish a unified optimization framework that captures the trade-off between communication throughput and sensing accuracy. We cast the problem as a weighted-sum scalarization under spatial deployment constraints.   
     
    \item 
    Given the non-convexity and coupled variables of the problem, we develop a {consensus ADMM}-based distributed algorithm, which decomposes the global optimization into local projection updates, proxy-assisted gradient-based consensus steps, and parallel dual variable updates. 
    The algorithm enables all UAVs to iteratively align their local decisions and reach consensus on the optimal swarm geometry, achieving global consistency without centralized coordination.  
		
	\item 
    We derive closed-form expressions for the gradients of both the achievable rate and CRB with respect to UAV positions, which allow each UAV to perform local gradient-based updates while maintaining consensus globally through limited information exchange. 
    This facilitates the efficient and scalable implementation of the proposed consensus ADMM algorithm.  
\end{itemize}

As corroborated by extensive simulations, the proposed consensus ADMM algorithm converges efficiently with high scalability in large-scale deployments of the UAV swarm-enabled ISAC system. Moreover, the UAV swarm-enabled ISAC framework achieves substantial performance gains over fixed-array baselines, exhibiting nearly an order-of-magnitude improvement in the joint communication–sensing metric.

The rest of this paper is organized as follows. Section~II introduces the system model. Section~III formulates the problem of UAV placement, characterizing the trade-off between communication and sensing. Section~IV presents the proposed optimization framework and discusses the scalarization approach for balancing the conflicting objectives. Section~V develops the proposed consensus ADMM-based distributed algorithm. Section~VI presents numerical results to verify the effectiveness and scalability of the proposed algorithm. Section~VII concludes the paper with future directions.

\textit{Notation:} 
$\mathbb{R}$ and $\mathbb{C}$ denote the sets of real and complex numbers, respectively; 
vectors appear as boldface lowercase letters (e.g., $\mathbf{x}$), and matrices as boldface uppercase letters (e.g., $\mathbf{A}$); 
for a vector $\mathbf{x}$, $x[i]$ denotes its $i$-th entry; 
for a matrix $\mathbf{A}$, $A_{ij}$ denotes its $(i,j)$-th element, $\lambda_i(\mathbf{A})$ denotes its $i$-th largest eigenvalue, and $\text{Diag}(\mathbf{A})$ denotes the diagonal matrix retaining its diagonal entries; $(\cdot)^\mathrm{\top}$, $(\cdot)^\mathrm{H}$, $(\cdot)^\ast$, and $\text{Tr}(\cdot)$ denote transpose, Hermitian transpose, complex conjugate, and trace, respectively; 
$|\cdot|_2$ denotes the Euclidean norm of a vector;
$|\cdot|_F$ denotes the Frobenius norm of a matrix; 
$\mathbf{I}$ denotes the identity matrix; 
$\mathbb{E}[\cdot]$ denotes expectation; $\odot$ stands for the Hadamard (element-wise) product; 
$\mathbb{I}_{\mathcal{S}}(\cdot)$ is the indicator function of a set $\mathcal{S}$, which takes $0$ if the argument lies in $\mathcal{S}$, or $+\infty$ otherwise.

\section{System Model: UAV Swarm-Enabled ISAC}

\subsection{UAV Swarm as a Virtual Antenna Array}

We consider a UAV swarm consisting of $N$ single-antenna unmanned aerial vehicles (UAVs), each equipped with an omnidirectional antenna and capable of maneuvering freely within a confined 3D space \cite{8344105}. These UAVs cooperatively form a virtual antenna array \cite{dohler2006distributed}, enabling spatial diversity and beamforming performance analogous to a conventional physical antenna array\footnote{The cooperation among UAVs relies on reliable and low-latency inter-UAV communication links (e.g., short-range wireless or fusion-node-assisted links), which ensure coherent signal processing, as in~\cite{5706317,7827017}.}.

Let $\mathbf{q}_i \in \mathbb{R}^3$ denote the 3D Cartesian coordinates of UAV $i$, $i = 1, \dots, N$. The swarm position matrix is defined as
\[
\mathbf{Q} = [\mathbf{q}_1, \dots, \mathbf{q}_N]^\top \in \mathbb{R}^{N \times 3}.
\]
The received signals of all UAVs are jointly processed to form beams for communication or sense the user’s locations. 

The position of each UAV is constrained within a maximum displacement radius $r_{\max}$ from its initial position $\mathbf{q}_{i,0}$, i.e.,
\[
\|\mathbf{q}_i - \mathbf{q}_{i,0}\| \leq r_{\max}, \quad \forall i.
\]

\subsection{Distributed MIMO Channel Modeling}

\begin{figure}[t]
	\centering
	\includegraphics[width=0.8\columnwidth]{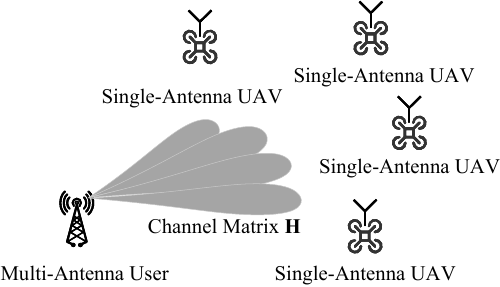}
	\caption{UAV swarm-enabled ISAC system model.}
	\label{sys_model}
\end{figure}

As illustrated in Fig.~\ref{sys_model}, a ground user equipped with $M$ transmit antennas communicates with a UAV swarm that collectively forms a distributed multiple-input multiple-output (MIMO) system. 
The user is located at an unknown position $\mathbf{p}_u \in \mathbb{R}^3$, and the $m$-th antenna of the user is positioned at $\mathbf{p}_u + \boldsymbol{\delta}_m$, where $\boldsymbol{\delta}_m \in \mathbb{R}^3$ denotes the offset vector of the $m$-th antenna from the array center.

We consider a narrowband line-of-sight (LoS) uplink communication scenario. Under the far-field assumption, the LoS channel coefficient between the $m$-th transmit antenna of the user and the $n$-th UAV is modeled as
\begin{equation}
	[\mathbf{H}]_{n,m} = \frac{\beta_0}{r_{n,m}^\gamma} 
	\cdot \exp\left( -j \frac{2\pi}{\lambda} r_{n,m} \right), \quad
	\mathbf{H} \in \mathbb{C}^{N \times M},
	\label{eq:channel_los}
\end{equation}
where $r_{n,m} = \|\mathbf{q}_n - (\mathbf{p}_u + \boldsymbol{\delta}_m)\|$ is the Euclidean distance between the $n$-th UAV and the $m$-th transmit antenna of the user;  
$\lambda$ denotes the carrier wavelength, 
$\gamma$ is the path-loss exponent (e.g., $\gamma = 2$ for free space), 
and $\beta_0$ represents the reference channel gain at unit distance.

\section{Formulation of UAV Swarm Placement}

In this section, we characterize the communication and sensing performance of the proposed UAV swarm-enabled ISAC system. 
The communication performance is evaluated by the achievable uplink rate from the user to the UAV swarm, 
while the sensing accuracy is quantified by the Cramér-Rao Bound (CRB) for user localization.

In the considered UAV swarm-enabled ISAC system, uplink communication is established from a multi-antenna ground user to the distributed UAVs.
The user transmits data streams through its $M$ antennas, while the $N$ spatially distributed UAVs jointly receive the signals, 
forming a virtual antenna array through coherent combining. 
This setup yields an equivalent uplink MIMO channel between the user and the UAV swarm. 
The achievable uplink capacity is given as follows.

\begin{proposition}\label{pro1}
	Assuming that the transmitted signal is spatially white Gaussian and the noise is additive white Gaussian with variance $\sigma^2$, the capacity of the uplink MIMO channel from the user to the UAV swarm is given by
	\begin{equation}\label{eq2}
		R = \log \det \big( \mathbf{I}_N + \frac{1}{\sigma^2} \mathbf{H} \mathbf{H}^H \big).
	\end{equation}
\end{proposition}

\noindent\textit{Proof:} See Appendix~\ref{appa}.

\smallskip
\smallskip

Proposition~\ref{pro1} characterizes the total spectral efficiency of the virtual MIMO system established between the user and the UAV swarm, which is inherently affected by the swarm geometry encoded in the channel matrix $\mathbf{H}$. 
In particular, well-separated UAVs can provide greater spatial diversity and improved channel rank conditions.

The channel capacity in \eqref{eq2} assumes that the channel state information (CSI) of the uplink MIMO channel is perfectly known at the receiver. 
In practice, such CSI can be obtained through uplink pilot-based training \cite{li2019hybrid,xie2016overview}, 
where the user transmits predefined pilot signals from all transmit antennas. 
Each UAV receives the pilots and estimates its corresponding uplink channel using standard techniques, such as least squares (LS) or minimum mean square error (MMSE) estimation. 
Thus, the UAV swarm has the necessary CSI for evaluating communication performance.

In the considered uplink ISAC framework, the UAV swarm performs receive-side sensing by reusing the user’s transmitted communication signals. 
Each UAV collects the user’s uplink signals and estimates their propagation delays, 
which serve as the sensing features for localization without requiring any dedicated sensing waveform. 
With this sensing mechanism, the sensing performance is characterized as follows.

\begin{proposition}
	When the user transmits known narrowband pilot signals and the measurement noise is Gaussian, 	the CRB for estimating the user’s position $\mathbf{p}_u \in \mathbb{R}^3$ based on the observed signal delays at the UAV swarm is given by
	\begin{equation}\label{crb}
		\mathrm{CRB}(\mathbf{p}_u) = \varsigma^2 c^2 \cdot \operatorname{Tr} \bigg( \Big( \sum_{n=1}^{N} \sum_{m=1}^{M} \mathbf{u}_{n,m} \mathbf{u}_{n,m}^\top \Big)^{-1} \bigg),
	\end{equation}
	where $\varsigma^2$ is the delay measurement noise variance, $c$ is the speed of light, 
	and $\mathbf{u}_{n,m}$ denotes the unit vector from the $m$-th antenna of the user to the $n$-th UAV, defined as 
	$\mathbf{u}_{n,m} = (\mathbf{q}_n - (\mathbf{p}_u + \boldsymbol{\delta}_m)) / \|\mathbf{q}_n - (\mathbf{p}_u + \boldsymbol{\delta}_m)\|$.
\end{proposition}

\noindent\textit{Proof:} See Appendix~B.

\smallskip
\smallskip

The CRB is inversely related to the Fisher Information Matrix (FIM) \cite{fisher1922mathematical}, 
which aggregates geometric information from all propagation paths between the user and the UAV swarm. 
The rank-one outer product $\mathbf{u}_{n,m} \mathbf{u}_{n,m}^\top$ captures the directional contribution of each propagation path between 
the $m$-th transmit antenna of the user and the $n$-th UAV.

The overall sensing performance is thus tightly coupled with the swarm geometry relative to the user. 
Specifically, the spatial configuration of the UAV swarm, i.e., the relative positions and spatial distribution of the UAVs, directly determines the directional diversity of the received signals. 
A well-spread UAV formation enhances the rank and conditioning of the FIM, thereby reducing the CRB and improving localization accuracy. 
Hence, the CRB serves as a tractable and geometry-aware metric to guide UAV positioning for high-precision sensing in ISAC systems.

Note that the communication and sensing performance are established under the assumption of full cooperation among the UAVs, 
which requires synchronization in both timing and phase. 
In practical implementations, such cooperation can be achieved by aggregating raw signals at a fusion node (e.g., a leader UAV or edge server) \cite{chepuri2012joint}, 
or through distributed consensus algorithms enabled by inter-UAV communication~\cite{11077738,10506083}. 
To strike a balance between communication throughput and sensing accuracy in the considered UAV swarm-enabled ISAC system, 
we formulate a joint optimization problem for the 3D placement of the UAVs. 
The communication metric is quantified by the achievable uplink rate $R(\mathbf{Q})$ in \eqref{eq2}, 
while the sensing performance is evaluated via the CRB for user localization $\mathrm{CRB}(\mathbf{Q})$ in \eqref{crb}. 
To jointly optimize these two objectives, we adopt a linear scalarization approach and formulate the following weighted-sum problem:    
\begin{equation}
	\begin{aligned}
		\min_{\mathbf{Q}} 
		\quad & - R(\mathbf{Q}) + \omega \cdot \mathrm{CRB}(\mathbf{Q}) \\
		\text{s.t.} \quad & \|\mathbf{q}_n - \mathbf{q}_{n,0}\|_2 \leq r_{\max}, \quad \forall n,
	\end{aligned}
	\label{eq:ISAC_opt}
\end{equation}
where $\omega > 0$ is a tunable weight reflecting the relative importance of sensing accuracy over communication rate.

Problem~\eqref{eq:ISAC_opt} aims to determine the UAV swarm geometry that achieves a desirable balance between communication efficiency and localization accuracy, while respecting flight constraints. 
By varying $\omega$, the entire Pareto frontier of the rate–sensing trade-off can be explored.

\begin{remark}
	In \eqref{eq:ISAC_opt}, both $R(\mathbf{Q})$ and $\mathrm{CRB}(\mathbf{Q})$ depend on the user position $\mathbf{p}_u$ through the channel model. 
	In practice, $\mathbf{p}_u$ is generally unknown and must be inferred from sensing measurements, creating a \textit{chicken–egg dilemma}: 
	optimizing the UAV positions requires knowledge of $\mathbf{p}_u$, while accurate localization depends on optimized UAV placement. 
	A practical solution is to adopt an iterative localization–optimization process, where an initial coarse position estimate is obtained and used for placement optimization, followed by iterative refinement based on the updated geometry until convergence.
\end{remark}

\section{Decentralized Consensus ADMM for UAV Swarm Placement}

In this section, we develop an algorithm to solve problem~\eqref{eq:ISAC_opt}. 
Traditional ISAC optimization methods require collecting the information of the entire network for centralized computation, 
which is impractical in the UAV swarm. 
Each UAV can only access its local information, such as local channel state information, and has limited computational capability.
To address this issue, we develop a distributed solution based on the consensus ADMM framework~\cite{boyd2011distributed,nishihara2015general,wang2019global} 
to solve problem~\eqref{eq:ISAC_opt}. 
This method allows each UAV to update its local variables via limited information exchange with neighboring UAVs, 
while ensuring consistency through consensus steps. 

\subsection{Consensus ADMM-Based Reformulation}

To enable decomposition, we introduce auxiliary consensus variables $\mathbf{z}_n \in \mathbb{R}^3$ for each UAV and rewrite problem \eqref{eq:ISAC_opt} as
\begin{equation}
	\begin{aligned}
		\min_{\{\mathbf{q}_n\}, \{\mathbf{z}_n\}} \quad & f(\{\mathbf{z}_n\}) := -R(\mathbf{Z}) + \omega \cdot \mathrm{CRB}(\mathbf{Z}) \\
		\text{s.t.} \quad & \mathbf{q}_n = \mathbf{z}_n, \quad \mathbf{q}_n \in \mathcal{Q}_n, \quad \forall n, \label{5}
	\end{aligned}
\end{equation}
where $\mathcal{Q}_n = \left\{ \mathbf{q} \in \mathbb{R}^3 : \|\mathbf{q} - \mathbf{q}_{n,0}\| \leq r_{\max} \right\}$ 
denotes the feasible flight region of the $n$-th UAV.

We convert the constraints $\mathbf{q}_n \in \mathcal{Q}_n$ into indicator functions, defined as
\[
\mathbb{I}_{\mathcal{Q}_n}(\mathbf{q}_n) = 
\begin{cases}
	0, & \text{if } \mathbf{q}_n \in \mathcal{Q}_n; \\
	+\infty, & \text{otherwise}.
\end{cases}
\]
Problem \eqref{5} can be expressed in the standard ADMM form:
\begin{equation}
	\min_{\{\mathbf{q}_n\}, \{\mathbf{z}_n\}} 
	\quad f(\{\mathbf{z}_n\}) + \sum_{n=1}^N \mathbb{I}_{\mathcal{Q}_n}(\mathbf{q}_n)
	\quad \text{s.t.} \quad \mathbf{q}_n = \mathbf{z}_n.
	\label{eq:consensus_admm}
\end{equation}
The augmented Lagrangian of \eqref{eq:consensus_admm} is given by
\begin{align}
	\mathcal{L}_\rho(\{\mathbf{q}_n\}, \{\mathbf{z}_n\}, &\{\boldsymbol{\mu}_n\}) 
	= f(\{\mathbf{z}_n\}) 
	+ \sum_{n=1}^N \Big[
	\mathbb{I}_{\mathcal{Q}_n}(\mathbf{q}_n)\nonumber
	 \\
	&+ \boldsymbol{\mu}_n^\top (\mathbf{q}_n - \mathbf{z}_n)  + \frac{\rho}{2} \|\mathbf{q}_n - \mathbf{z}_n\|^2 
	\Big],\label{7}
\end{align}
where $\boldsymbol{\mu}_n, \forall n$, are the dual variables.

\subsection{Consensus ADMM Iteration}

The consensus ADMM algorithm proceeds by iteratively updating the primal variables $\{\mathbf{q}_n\}$ and $\{\mathbf{z}_n\}$, 
and the dual variables $\{\boldsymbol{\mu}_n\}$. Here, the superscript $(\cdot)^k$ denotes the $k$-th iteration of consensus ADMM. 
The details follow.

\paragraph{Step 1: Local Variable Update ($\{\mathbf{q}_n\}$)}

At each iteration $k+1$, each UAV updates its local variable $\mathbf{q}_n$ by minimizing $\mathcal{L}_\rho$ 
while fixing $\mathbf{z}_n = \mathbf{z}_n^k$ and $\boldsymbol{\mu}_n = \boldsymbol{\mu}_n^k$:
\begin{equation}
	\mathbf{q}_n^{k+1} = \arg\min_{\mathbf{q}_n} \ 
	\mathbb{I}_{\mathcal{Q}_n}(\mathbf{q}_n) 
	+ {(\boldsymbol{\mu}_n^k)}^{\top} (\mathbf{q}_n - \mathbf{z}_n^k)
	+ \frac{\rho}{2} \|\mathbf{q}_n - \mathbf{z}_n^k\|^2.
	\label{eq:qn_raw}
\end{equation}
We reorganize the second and third terms on the right-hand side (RHS) of \eqref{eq:qn_raw} as
\begin{align}
	\|\mathbf{q}_n - \mathbf{z}_n^k\|^2 
	&+ \frac{2}{\rho} {(\boldsymbol{\mu}_n^k)}^{\top} (\mathbf{q}_n - \mathbf{z}_n^k) \nonumber \\
	&= \left\| \mathbf{q}_n - \mathbf{z}_n^k 
	+ \frac{1}{\rho} \boldsymbol{\mu}_n^k \right\|^2 
	- \frac{1}{\rho^2} \|\boldsymbol{\mu}_n^k\|^2.
\end{align}
Thus, \eqref{eq:qn_raw} can be rewritten as
\begin{align}
	\mathbf{q}_n^{k+1} 
	&= \arg\min_{\mathbf{q}_n} \ 
	\mathbb{I}_{\mathcal{Q}_n}(\mathbf{q}_n) \nonumber \\
	&\quad 
	+ \frac{\rho}{2} \left\| \mathbf{q}_n - \mathbf{z}_n^k 
	+ \frac{1}{\rho} \boldsymbol{\mu}_n^k \right\|^2 
	- \frac{1}{2\rho} \|\boldsymbol{\mu}_n^k\|^2.
\end{align}

Since $\frac{1}{2\rho} \|\boldsymbol{\mu}_n^k\|^2$ is independent of $\mathbf{q}_n$, 
the update of $\mathbf{q}_n$ becomes a projection problem given by
\begin{equation}
		\mathbf{q}_n^{k+1} = \arg\min_{\mathbf{q}_n \in \mathcal{Q}_n} 
		\left\| \mathbf{q}_n - \left( \mathbf{z}_n^k - \frac{1}{\rho} \boldsymbol{\mu}_n^k \right) \right\|^2,
\end{equation}
which leads to the closed-form solution:
\begin{equation}
		\mathbf{q}_n^{k+1} = \Pi_{\mathcal{Q}_n} \left( \mathbf{z}_n^k - \frac{1}{\rho} \boldsymbol{\mu}_n^k \right),
\end{equation}
where $\Pi_{\mathcal{Q}_n}(\cdot)$ denotes the Euclidean projection onto the feasible set 
$\mathcal{Q}_n = \left\{ \mathbf{q} \in \mathbb{R}^3 : \|\mathbf{q} - \mathbf{q}_{n,0}\|_2 \leq r_{\max} \right\}$,
representing the allowable flight region of UAV $n$.

We now derive the closed-form expression for projecting a point $\mathbf{a}=\mathbf{z}_n^k - \frac{1}{\rho} \boldsymbol{\mu}_n^k  \in \mathbb{R}^3$ 
onto a Euclidean ball centered at $\mathbf{q}_{n,0}$ with radius $r_{\max}$. 
That is, the feasible set is
\begin{equation}
	\mathcal{Q}_n = \left\{ \mathbf{q}_n \in \mathbb{R}^3 : \|\mathbf{q}_n - \mathbf{q}_{n,0}\|_2 \leq r_{\max} \right\}.
\end{equation}
The projection of $\mathbf{a}\in \mathbb{R}^3$ on $\mathcal{Q}_n$ is obtained from
\begin{equation}
	\Pi_{\mathcal{Q}_n}(\mathbf{a}) := \arg\min_{\mathbf{q}_n \in \mathcal{Q}_n} \|\mathbf{q}_n - \mathbf{a}\|_2^2. \label{14}
\end{equation}
We consider two cases:

{\emph{Case 1}: $\mathbf{a} \in \mathcal{Q}_n$}.  
If $\mathbf{a}$ lies within the feasible region, i.e.,
\begin{equation}
	\|\mathbf{a} - \mathbf{q}_{n,0}\|_2 \leq r_{\max},
\end{equation}
then the optimal solution to \eqref{14} is  $\mathbf{a}$; i.e.,
\begin{equation}
	\Pi_{\mathcal{Q}_n}(\mathbf{a}) = \mathbf{a}.
\end{equation}

{\emph{Case 2}: $\mathbf{a} \notin \mathcal{Q}_n$}.  
If $\mathbf{a}$ lies outside the feasible region, the projection is the unique point on the sphere boundary along the line segment connecting $\mathbf{q}_{n,0}$ and $\mathbf{a}$. 
Define the direction vector:
\begin{equation}
	\mathbf{d} := \mathbf{a} - \mathbf{q}_{n,0}.
\end{equation}
We seek a scalar $\alpha > 0$ such that the projected point satisfies
\begin{equation}
	\mathbf{q}_n^\star = \mathbf{q}_{n,0} + \alpha \mathbf{d}, \label{18}
\end{equation}
and lies on the boundary:
\begin{equation}
	\|\mathbf{q}_n^\star - \mathbf{q}_{n,0}\|_2 = r_{\max}. \label{19}
\end{equation}
Substituting \eqref{18} into \eqref{19} yields
\begin{equation}
	\alpha = \frac{r_{\max}}{\|\mathbf{d}\|_2}.
\end{equation}
Thus, the projection point is
\begin{equation}
	\Pi_{\mathcal{Q}_n}(\mathbf{a}) = \mathbf{q}_{n,0} + \frac{r_{\max}}{\|\mathbf{a} - \mathbf{q}_{n,0}\|_2} (\mathbf{a} - \mathbf{q}_{n,0}).
\end{equation}

Combining \textit{Case~1} and \textit{Case~2}, we obtain the final closed-form projection expression:
\begin{equation}
	\Pi_{\mathcal{Q}_n}(\mathbf{a}) = 
	\begin{cases}
		\mathbf{a}, & \!\!\!\!\!\!\!\!\!\!\!\!\!\!\!\!\!\!\!\!\!\!\!\!\!\!\!\!\!\text{if } \|\mathbf{a} - \mathbf{q}_{n,0}\|_2 \leq r_{\max}; \\
		\mathbf{q}_{n,0} 
		+ \dfrac{r_{\max}}{\|\mathbf{a} - \mathbf{q}_{n,0}\|_2}
		(\mathbf{a} - \mathbf{q}_{n,0}), & \text{otherwise}.
	\end{cases}\label{22}
\end{equation}
This expression is applied in Step~1 of the consensus ADMM algorithm for updating the local position variable $\mathbf{q}_n$.

\begin{remark} 
	In the consensus ADMM framework, each UAV \(n\) maintains its own local variables \(\{\mathbf{q}_n^k, \mathbf{z}_n^k, \boldsymbol{\mu}_n^k\}\).  
	The local update in Step~1 follows
	\[
	\mathbf{q}_n^{k+1} = \Pi_{\mathcal{Q}_n}\!\left( \mathbf{z}_n^k - \tfrac{1}{\rho}\boldsymbol{\mu}_n^k \right).
	\]
	This update depends only on UAV \(n\)’s own state variables and feasible flight region, 
	and can be executed locally without any information exchange with other UAVs in the swarm.  
\end{remark}

\paragraph{Step 2: Consensus Variable Update ($\{\mathbf{z}_n\}$)}

We next derive the update rule for the consensus variables $\{\mathbf{z}_n\}$ in the ADMM procedure.
At the $(k+1)$-th iteration of consensus ADMM, given the current estimates $\{\mathbf{q}_n^{k+1}\}$ and $\{\boldsymbol{\mu}_n^k\}$, the consensus variables are updated by minimizing the augmented Lagrangian with respect to $\{\mathbf{z}_n\}$:
\[
\{\mathbf{z}_n^{k+1}\} = \arg\min_{\{\mathbf{z}_n\}} \ \mathcal{L}_\rho(\{\mathbf{q}_n^{k+1}\}, \{\mathbf{z}_n\}, \{\boldsymbol{\mu}_n^k\}).
\]
Extracting the terms that depend on $\{\mathbf{z}_n\}$ yields
\begin{align}
	\{\mathbf{z}_n^{k+1}\} 
	&= \arg\min_{\{\mathbf{z}_n\}} \ 
	f(\{\mathbf{z}_n\}) 
	- \sum_{n=1}^N {(\boldsymbol{\mu}_n^k)}^{\top} \mathbf{z}_n \notag \\
	&\quad
	+ \frac{\rho}{2} \sum_{n=1}^N 
	\|\mathbf{q}_n^{k+1} - \mathbf{z}_n\|^2 .\label{23}
\end{align}
We rearrange the second and third terms on the RHS of \eqref{23}:
\begin{align}
	&\|\mathbf{q}_n^{k+1} - \mathbf{z}_n\|^2 
	- \frac{2}{\rho} {(\boldsymbol{\mu}_n^k)}^{\top} \mathbf{z}_n \notag \\
	&\quad= \left\| \mathbf{q}_n^{k+1} - \mathbf{z}_n 
	+ \frac{1}{\rho} \boldsymbol{\mu}_n^k \right\|^2
	- \frac{1}{\rho^2} \|\boldsymbol{\mu}_n^k\|^2, \ \forall n,
\end{align}
where $\frac{1}{\rho^2} \|\boldsymbol{\mu}_n^k\|^2$ is independent of $\mathbf{z}_n$. As a result, the consensus variables $\{\mathbf{z}_n\}$ are updated by solving
\begin{equation}
	\mathbf{Z}^{k+1} 
	= \arg\min_{\mathbf{Z}} \ f(\mathbf{Z})
	+ \frac{\rho}{2} \sum_{n=1}^N 
	\left\| \mathbf{q}_n^{k+1} - \mathbf{z}_n + \frac{1}{\rho} \boldsymbol{\mu}_n^k \right\|^2,
	\label{eq:z_update_step}
\end{equation}
where $\mathbf{Z} \triangleq [\mathbf{z}_1,  \ldots, \mathbf{z}_N]$ stacks all consensus variables by rows.
This step minimizes the smooth global function $f(\mathbf{Z}) = -R(\mathbf{Z}) + \omega \cdot \mathrm{CRB}(\mathbf{Z})$ regularized by a separable quadratic term without constraints; in practice, it can be solved via gradient descent \cite{shor2012minimization}, depending on the structure of $f$ in \eqref{5}.

\begin{proposition}\label{pro3}
The gradient of the objective in \eqref{eq:z_update_step} with respect to the $n$-th UAV’s consensus variable is given by
\begin{align}
	&\nabla_{\mathbf{z}_n} \Big( f(\mathbf{Z}) + \tfrac{\rho}{2}
	\sum_{i=1}^N \| \mathbf{q}_i^{k+1} - \mathbf{z}_i 
	+ \tfrac{1}{\rho}\boldsymbol{\mu}_i^k \|^2 \Big) \nonumber \\
	&= - \nabla_{\mathbf{z}_n} \!R(\mathbf{Z})
	\!+\! \omega \nabla_{\mathbf{z}_n} \mathrm{CRB}(\mathbf{Z})
	\!+\! \rho \mathbf{z}_n \!-\! \rho \mathbf{q}_n^{k+1} \!-\! \boldsymbol{\mu}_n^k ;
	\label{eq:grad_objective_final}
\end{align}
the gradient of the communication rate with respect to the UAV position is given by
\begin{align}
	\nabla_{\mathbf{z}_n} R(\mathbf{Z}) 
	&=\!\! \sum_{m=1}^M \!\!\Re \Bigg\{ 
	\Big( \tfrac{2}{\sigma^2}[\mathbf{A}^{-1}\mathbf{H}]_{n,m} \Big)^{\!*}
	\beta_0 \Big( \!\!-\tfrac{\gamma}{r_{n,m}^{\gamma+2}} 
	- j \tfrac{2\pi}{\lambda}\tfrac{1}{r_{n,m}^{\gamma+1}} \Big) \nonumber \\
	&\quad \times e^{-j\frac{2\pi}{\lambda} r_{n,m}} 
	\mathbf{d}_{n,m} \Bigg\}.
	\label{eq:grad_rate_final}
\end{align}
Consequently, the gradient of the CRB with respect to the UAV position is given by
\begin{equation}
	\nabla_{\mathbf{z}_n}\!\mathrm{CRB}(\mathbf{Z}) \!=\!
	-\varsigma^2 \!c^2 \!\!
	\begin{bmatrix}
		\mathrm{Tr}\!\Big(\mathbf{J}^{-1}\!\sum_{m=1}^M\!\! \frac{\partial (\mathbf{u}_{n,m}\mathbf{u}_{n,m}^\top)}{\partial (\mathbf{z}_n)_1} \mathbf{J}^{-1}\Big) \\
		\mathrm{Tr}\!\Big(\mathbf{J}^{-1}\!\sum_{m=1}^M \!\!\frac{\partial (\mathbf{u}_{n,m}\mathbf{u}_{n,m}^\top)}{\partial (\mathbf{z}_n)_2} \mathbf{J}^{-1}\Big) \\
		\mathrm{Tr}\!\Big(\mathbf{J}^{-1}\!\sum_{m=1}^M \!\!\frac{\partial (\mathbf{u}_{n,m}\mathbf{u}_{n,m}^\top)}{\partial (\mathbf{z}_n)_3} \mathbf{J}^{-1}\Big)
	\end{bmatrix}.\label{28}
\end{equation}
Here, $\mathbf{d}_{n,m}=\mathbf{z}_n-(\mathbf{p}_u+\boldsymbol{\delta}_m)$,
$r_{n,m}=\|\mathbf{d}_{n,m}\|$, 
$\mathbf{A}=\mathbf{I}_N+\tfrac{1}{\sigma^2}\mathbf{H}\mathbf{H}^H,$ 
$\mathbf{J} 	= \sum_{i=1}^N \sum_{m=1}^M \mathbf{u}_{i,m}\mathbf{u}_{i,m}^\top,$ and
$
\frac{\partial (\mathbf{u}_{n,m}\mathbf{u}_{n,m}^\top)}{\partial (\mathbf{z}_n)_k}
= \frac{1}{r_{n,m}}
\Big[(\mathbf{I}-\mathbf{u}_{n,m}\mathbf{u}_{n,m}^\top)\mathbf{e}_k \mathbf{u}_{n,m}^\top 
+ \mathbf{u}_{n,m} \mathbf{e}_k^\top (\mathbf{I}-\mathbf{u}_{n,m}\mathbf{u}_{n,m}^\top)\Big].
$
\end{proposition}
\noindent\textit{Proof:} See Appendix~\ref{app_c}.

\smallskip
\smallskip

Based on Proposition~\ref{pro3}, we obtain the gradient of the objective function in \eqref{eq:z_update_step} with respect to each consensus variable \(\mathbf{z}_n\). Consequently, Step~2 can be carried out using a gradient descent procedure to update \(\{\mathbf{z}_n^{k+1}\}\). 

\begin{remark}  
 	The gradient expression in \eqref{eq:grad_objective_final} involves the terms  
 	\(- \nabla_{\mathbf{z}_n} R(\mathbf{Z}) + \omega \nabla_{\mathbf{z}_n} \mathrm{CRB}(\mathbf{Z})\),  
 	whose evaluation requires global channel and geometry information. 
 	To enable a distributed implementation in a UAV swarm, this computation can be delegated to a {proxy UAV} (or edge server): 
 	each UAV \(n\) transmits its local consensus variable \(\mathbf{z}_n^k\) and dual variable \(\boldsymbol{\mu}_n^k\) to the proxy UAV, 
 	which then evaluates the gradients with respect to \(\mathbf{z}_n, \forall n\), and returns the corresponding components to the individual UAVs. 
 	After receiving its gradient, each UAV \(n\) performs the gradient descent update of \(\mathbf{z}_n\) locally.  	
 	In this way, only the gradient calculation requires centralized assistance; the gradient descent updates remain fully distributed across the UAV swarm. 
 	The resulting procedure resembles {inexact ADMM} \cite{chang2014multi,bai2022inexact}, 
 	where each subproblem is approximately solved by a few gradient steps, significantly reducing computational and communication overhead.
\end{remark}

\paragraph{Step 3: Dual Variable Update ($\{\boldsymbol{\mu}_n\}$)}
The dual variables are updated via a gradient ascent step:
\begin{equation}
	\boldsymbol{\mu}_n^{k+1} = \boldsymbol{\mu}_n^k + \rho \left( \mathbf{q}_n^{k+1} - \mathbf{z}_n^{k+1} \right).\label{29}
\end{equation}
This update depends only on the local variables \(\mathbf{q}_n^{k+1}\), \(\mathbf{z}_n^{k+1}\), and \(\boldsymbol{\mu}_n^k\) of UAV \(n\).  
Step 3 can be executed independently by all UAVs in parallel with no inter-UAV communication.  

\subsection{Summary of Proposed Consensus ADMM Algorithm}
Algorithm~\ref{alg:admm} summarizes the proposed consensus ADMM-based optimization framework for the UAV swarm-enabled ISAC system.  
The algorithm alternates between three updates.  
Step 1 projects each UAV’s position onto its feasible flight region and is fully parallelizable without communication.  
Step~2 updates the consensus variables using the gradients of the global objective, which require global information. To enable distributed operation, a {proxy UAV} (or edge server) computes these gradients and broadcasts them back to the swarm, after which each UAV locally updates its own $\mathbf{z}_n$.  
Step 3 performs the dual update, which relies solely on local information and is therefore decentralized.  
Consequently, only the gradient evaluation in Step 2 requires proxy assistance, while the main updates remain fully distributed.

 We prove the convergence of Algorithm~\ref{alg:admm} by analyzing the non-increasing property of the augmented Lagrangian function and the gradual reduction of the primal and dual residuals.

\subsubsection{Non-increasing Property of the Augmented Lagrangian}\label{sec: non-increasing property}

The augmented Lagrangian function used in Algorithm~\ref{alg:admm} is defined in \eqref{7}.
The algorithm updates the variables in three steps: 
In Step 1, each UAV updates its local variable \( \mathbf{q}_n \) according to \eqref{22} to minimize the augmented Lagrangian with respect to \( \mathbf{q}_n \). This ensures that \( \mathbf{q}_n \) is projected onto its feasible set \( Q_n \) and converges to a solution that minimizes the augmented Lagrangian. In Step~2, the consensus variables \( \mathbf{z}_n \) are updated by minimizing the augmented Lagrangian with respect to \( \mathbf{z}_n \) with gradient descent; see Proposition~\ref{pro3}. This step ensures that all UAVs reach consensus on the shared variables \( \mathbf{z}_n \). In Step 3, the dual variables \( \boldsymbol{\mu}_n \) are updated in \eqref{29} to enforce the constraints between \( \mathbf{q}_n \) and \( \mathbf{z}_n \). 
This step ensures that the primal and dual variables are aligned, which helps reduce constraint violations and guides the algorithm towards convergence.
Through these steps, the augmented Lagrangian function does not increase in each iteration; i.e., it is non-increasing and thus the algorithm progressively converges towards an optimal solution.

\subsubsection{Primal and Dual Residuals}\label{sec: primal and dual}
The primal residual is defined as the difference between the local variables \( \mathbf{q}_n \) and the consensus variables \( \mathbf{z}_n \), i.e., \( \| \mathbf{q}_n^{k+1} - \mathbf{z}_n^{k+1} \| \). As the iterations proceed, this residual decreases, and the UAVs' positions and consensus variables approach their optimal values. On the other hand, the dual residual is the difference between the updates of the dual variables \( \boldsymbol{\mu}_n \), i.e., \( \| \boldsymbol{\mu}_n^{k+1} - \boldsymbol{\mu}_n^k \| \). In each iteration, the dual residual decreases, ensuring the alignment of primal and dual variables.

The primal and dual residuals are expected to converge to zero as the number of iterations increases, implying that Algorithm~\ref{alg:admm} converges to a solution where both the primal and dual variables satisfy the optimality conditions.

\subsubsection{Convergence Proof}

According to Section \ref{sec: non-increasing property} and the standard convergence theory of ADMM \cite{Boyd2011ADMM}, we can conclude that the value of the augmented Lagrangian function decreases or remains unchanged in each iteration, i.e.,
\begin{align}
    L_{\rho}(\{\mathbf{q}_n^{k+1}\}, \{\mathbf{z}_n^{k+1}\}, \{\boldsymbol{\mu}_n^{k+1}\}) \!\leq\! L_{\rho}(\{\mathbf{q}_n^k\}, \{\mathbf{z}_n^k\}, \{\boldsymbol{\mu}_n^k\}).\!
\end{align}
To this end, Algorithm~\ref{alg:admm} moves towards a minimum. Moreover, according to Section~\ref{sec: primal and dual}, the primal residual \( \| \mathbf{q}_n^{k+1} - \mathbf{z}_n^{k+1} \| \) and the dual residual \( \| \boldsymbol{\mu}_n^{k+1} - \boldsymbol{\mu}_n^k \| \) decrease and converge to zero as the iterations progress. Any limit point of the sequence generated by Algorithm~\ref{alg:admm} can satisfy the first-order optimality conditions 
for the reformulated problem \eqref{eq:consensus_admm}, ensuring that the algorithm converges to a stationary point or optimal solution~\cite{Chang2014TSP,Nishihara2015ICML,Wang2019JSC}.
The convergence of Algorithm~\ref{alg:admm} is also numerically validated in Section \ref{sec: simulation}.

\begin{remark}
Algorithm~\ref{alg:admm} extends the standard ADMM to a consensus form tailored for distributed UAV swarm optimization.
Unlike the centralized ADMM, which alternates between globally coupled primal and dual updates, the proposed consensus ADMM decomposes the problem across UAVs:  
each UAV performs a local projection onto its feasible region, while the global objective is optimized through inexact gradient-based consensus steps requiring only lightweight coordination. The per-iteration complexity is $\mathcal{O}\!\left(N^3 + N^2M\right)$, offering scalability to moderately large UAV swarms.  
\end{remark}
\begin{remark}
{ Algorithm~\ref{alg:admm} enables the UAVs to reach consensus on a globally optimal swarm geometry.}
In each iteration, every UAV $n$ maintains two local states: the projected position $\mathbf{q}_n$ within its feasible flight region and the consensus variable $\mathbf{z}_n$ that represents the UAV’s local estimate of the globally agreed swarm geometry. 
Through iterative updates, the UAVs reconcile their local estimates $\mathbf{q}_n$ with the shared consensus variables $\mathbf{z}_n$, which are jointly refined via the proxy-assisted gradient step. 
{This process drives the swarm toward a common configuration that aligns all UAVs' local decisions, achieving geometric consensus.}
In practice, the consensus step corresponds to lightweight inter-UAV message exchanges or proxy coordination, ensuring all UAVs share a coherent understanding of the optimal swarm formation.
\end{remark}

\begin{algorithm}[t]
	\caption{Consensus ADMM for UAV swarm-based ISAC}
	\label{alg:admm}
	\begin{algorithmic}[1]
		\STATE \textbf{Initialize:} 
		UAV positions $\{\mathbf{q}_n^0\}_{n=1}^N$, consensus variables $\{\mathbf{z}_n^0\}_{n=1}^N$, dual variables $\{\boldsymbol{\mu}_n^0\}_{n=1}^N$, penalty parameter $\rho>0$, and step size $\eta$ for gradient descent.
		
		\FOR{$k=0,1,2,\ldots$ until convergence}
		\STATE \textbf{Step 1: Local variable update (fully distributed)} \\  
		Each UAV \(n\) updates
		\[
		\mathbf{q}_n^{k+1} = \Pi_{\mathcal{Q}_n}\!\left( \mathbf{z}_n^k - \tfrac{1}{\rho}\boldsymbol{\mu}_n^k \right),
		\]
		where $\Pi_{\mathcal{Q}_n}(\cdot)$ denotes projection onto the feasible flight region $\mathcal{Q}_n$.
		
		\STATE \textbf{Step 2: Consensus variable update (proxy-assisted)} \\  
		$a$. A proxy UAV collects $\{\mathbf{z}_n^k, \boldsymbol{\mu}_n^k\}_{n=1}^N$ and evaluates
		\[
		\nabla_{\mathbf{z}_n}\!\Big(f(\mathbf{Z}) + \tfrac{\rho}{2}\!\sum_{i=1}^N \|\mathbf{q}_i^{k+1} - \mathbf{z}_i + \tfrac{1}{\rho}\boldsymbol{\mu}_i^k\|^2\Big).
		\]  
		$b$. The computed gradients are broadcast to all UAVs. \\  
		$c$. Each UAV \(n\) updates
		\[
		\mathbf{z}_n^{k+1} = \mathbf{z}_n^k - \eta \, \nabla_{\mathbf{z}_n}(\cdot).
		\]		
		
		\STATE \textbf{Step 3: Dual variable update (fully distributed)} \\  
		Each UAV \(n\) updates
		\[
		\boldsymbol{\mu}_n^{k+1} = \boldsymbol{\mu}_n^k + \rho \left( \mathbf{q}_n^{k+1} - \mathbf{z}_n^{k+1} \right).
		\]
		\ENDFOR
	\end{algorithmic}
\end{algorithm}

\section{Numerical Results}\label{sec: simulation}
This section evaluates the proposed consensus ADMM-based framework for UAV swarm-enabled ISAC systems. All simulations were conducted in MATLAB R2021b. 

\subsection{Simulation Environment}
Unless otherwise specified, the default simulation parameters are set as follows.  
The UAV swarm consists of $N=4$ single-antenna UAVs that collaboratively form a distributed virtual antenna array.  
The ground user is equipped with $M=4$ transmit antennas.  
The carrier wavelength is $\lambda_c = 0.1$~m, and the reference channel gain is $\beta_0 = 1$ with a path-loss exponent $\gamma = 2$ \cite{rappaport2010wireless}.  
The noise variance is $\sigma^2 = 10^{-12}$ \cite{Goldsmith2005Book}, and the observation delay variance is $\varsigma^2 = 10^{-15}$.  
The maximum allowable movement radius for each UAV is $r_{\max} = 20$~m.  
The weighting parameter between communication and sensing objectives is $\omega = 1$.  
The speed of light is $c = 3\times 10^8$~m/s.  

The user is located at the origin $(0,0,0)$, and the initial UAV positions are randomly generated within a $100\times100\times100$~m$^3$ cube.  
The transmit antennas’ offsets $\{\boldsymbol{\delta}_m\}$ are drawn from a uniform distribution in $[-0.5,0.5]^3$~m.  
For the consensus ADMM algorithm, the penalty parameter is $\rho = 1$, the maximum number of iterations is $10^5$, and the step size of the gradient descent is $\eta = 10^{-3}$.

\subsection{Performance Evaluation under Different Settings}

\begin{figure}[t]
	\centering
	\includegraphics[width=0.85\columnwidth]{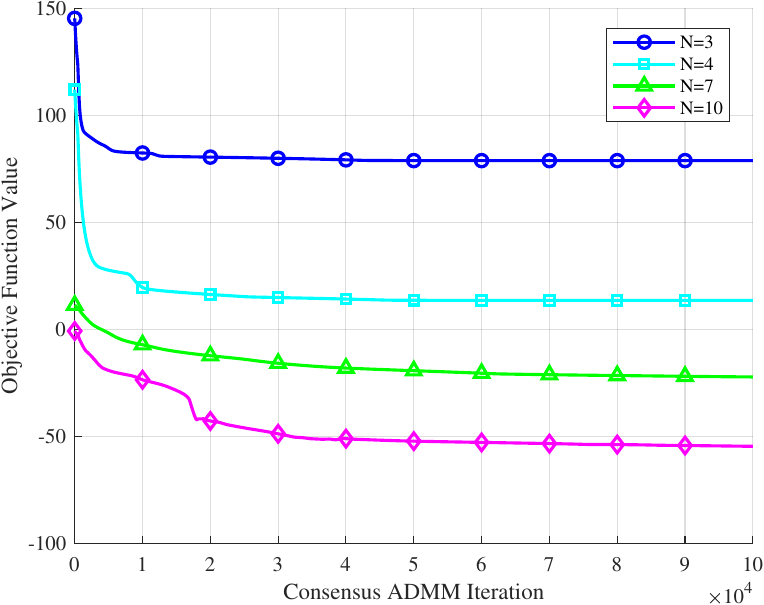}
	\caption{Convergence of the objective function under different UAV swarm sizes.}
	\label{fig:convergence}
\end{figure}

Fig.~\ref{fig:convergence} illustrates the convergence behavior of the proposed consensus ADMM algorithm for different swarm sizes $N=\{3,4,7,10\}$. It is observed that the objective function value decreases rapidly during the initial iterations and gradually stabilizes, demonstrating the convergence of the distributed optimization framework. The steady-state objective value decreases as the swarm size increases, indicating that deploying more UAVs provides greater spatial flexibility to balance communication and sensing performance.

\begin{figure}[t]
	\centering
	\subfloat[$N=3$]{%
		\includegraphics[width=0.48\columnwidth]{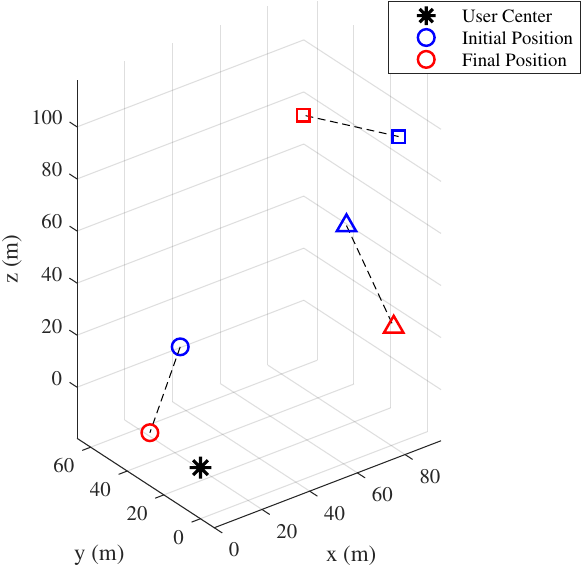}
	}
	\hfill
	\subfloat[$N=4$]{%
		\includegraphics[width=0.48\columnwidth]{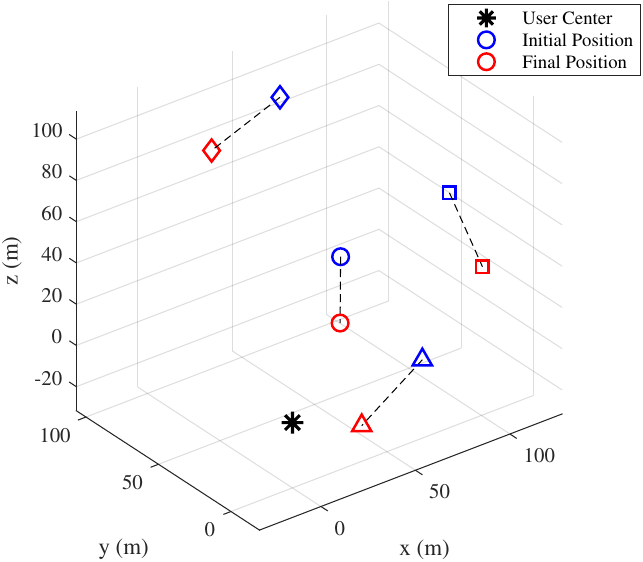}
	}
	
	\vspace{0.5em}
	
	\subfloat[$N=7$]{%
		\includegraphics[width=0.48\columnwidth]{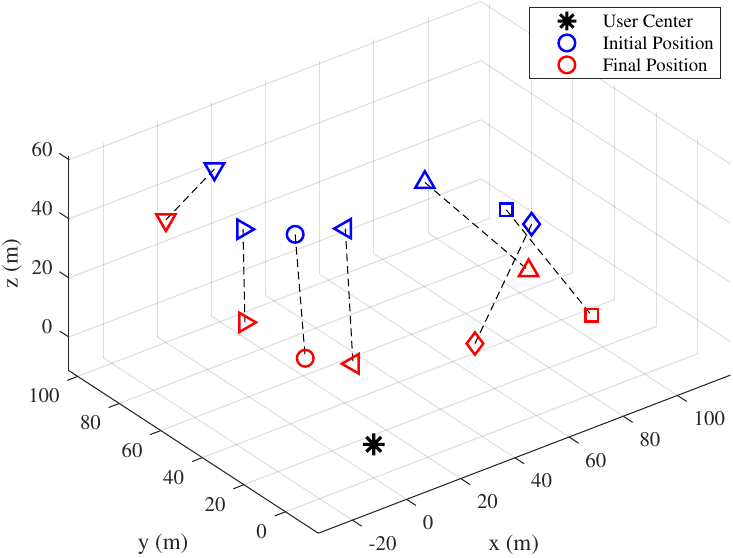}
	}
	\hfill
	\subfloat[$N=10$]{%
		\includegraphics[width=0.48\columnwidth]{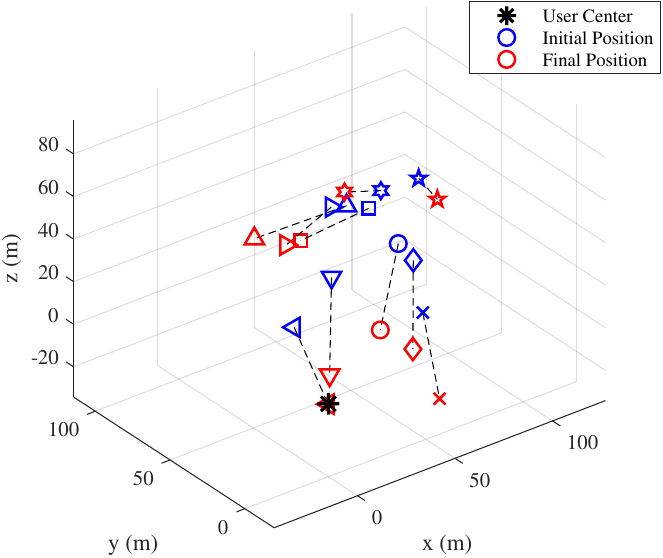}
	}
	
	\caption{UAV position updates (initial vs. optimized) for different swarm sizes.}
	\label{fig:positions}
\end{figure}

Fig.~\ref{fig:positions} shows the trajectory evolution of the UAVs under different swarm sizes. The user is located at the origin. The blue markers denote the initial UAV positions, while the red markers indicate their optimized final positions. It can be seen that the UAVs move significantly from their initial random deployment to more favorable configurations guided by the proposed consensus ADMM algorithm. As the swarm size $N$ increases, the UAVs become more spatially distributed around the user, forming an optimized 3D geometry that enhances communication capacity while reducing the localization CRB.

\begin{figure}[t]
	\centering
	\includegraphics[width=0.85\columnwidth]{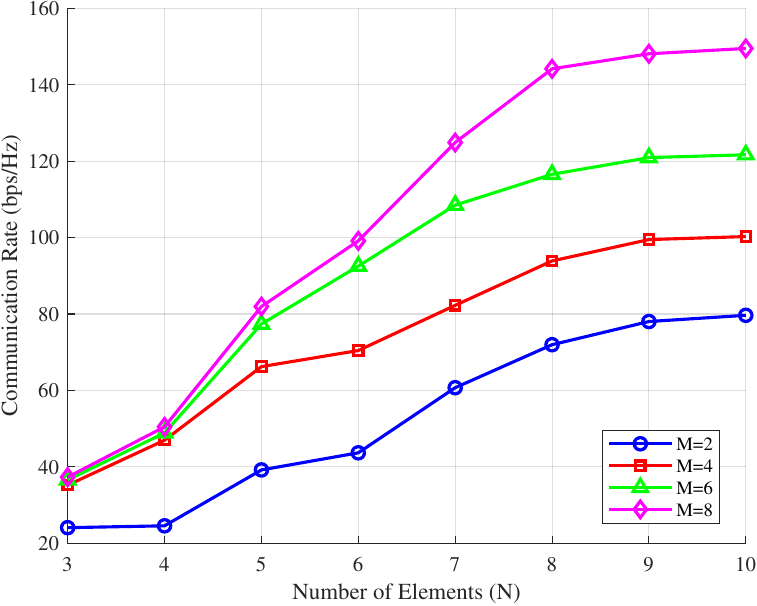}
	\caption{Achievable rate versus swarm size $N$ under different numbers of transmit antennas $M$ at the user.}
	\label{fig:diff_M_rate}
\end{figure}

\begin{figure}[t]
	\centering
	\includegraphics[width=0.85\columnwidth]{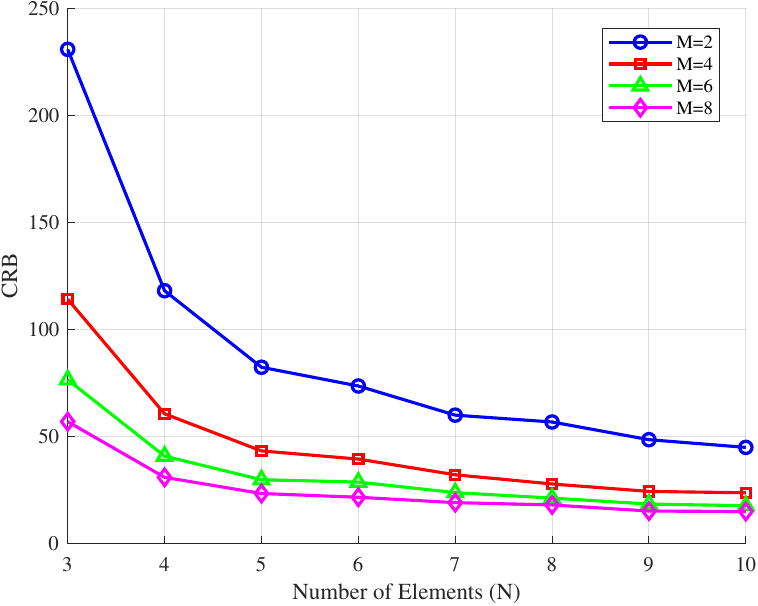}
	\caption{CRB performance versus swarm size $N$ under different numbers of transmit antennas $M$ at the user.}
	\label{fig:diff_M_CRB}
\end{figure}

We further evaluate the impact of the number $M$ of transmit antennas at the user side on the rate–sensing performance of the proposed UAV swarm-enabled ISAC framework.  
Fig.~\ref{fig:diff_M_rate} illustrates the achievable uplink rate as a function of the swarm size $N$ for different values of $M$, while Fig.~\ref{fig:diff_M_CRB} presents the corresponding CRB performance.  
As shown in Fig.~\ref{fig:diff_M_rate}, larger $N$ and/or $M$ yield notable rate improvements. Increasing $N$ expands the effective aperture of the distributed array, enabling stronger spatial multiplexing, while increasing $M$ provides more transmit degrees of freedom, thereby enhancing the rank of the channel matrix $\mathbf{H}$.  
These results demonstrate the synergistic benefits of combining multi-antenna transmitters with a distributed UAV swarm for ISAC.

In Fig.~\ref{fig:diff_M_CRB}, the CRB decreases monotonically with $N$, since a larger swarm provides richer geometric diversity and improves the conditioning of the FIM.  
Moreover, increasing $M$ also reduces the CRB for all $N$, owing to additional independent propagation paths that enhance spatial resolution.  
This confirms that a larger transmit antenna array significantly improves localization accuracy.

Overall, the results highlight that both communication and sensing performance can be substantially enhanced by increasing the number of user transmit antennas as well as the UAV swarm size.  
This observation suggests that the UAV swarm-enabled ISAC framework naturally complements multi-antenna transmitters, offering a scalable solution to improve the rate–sensing trade-off in future wireless networks.

To further investigate the communication–sensing trade-off, we examine system performance under different values of the weight parameter $\omega$ in problem~\eqref{eq:ISAC_opt}.  
Recall that $\omega$ controls the balance between minimizing the CRB and maximizing the communication rate.  
By varying $\omega$, we can effectively trace the Pareto frontier \cite{van2009probing} between these two conflicting objectives.

\begin{figure}[t]
	\centering
	\includegraphics[width=0.85\columnwidth]{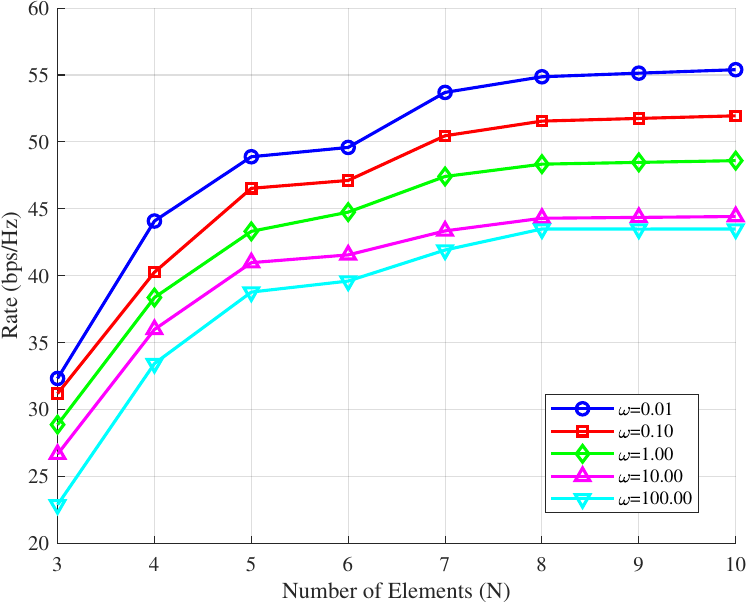}
	\caption{Communication rate performance versus swarm size $N$ under different values of the weight parameter $\omega$.}
	\label{fig:omega_rate}
\end{figure}

\begin{figure}[t]
	\centering
	\includegraphics[width=0.85\columnwidth]{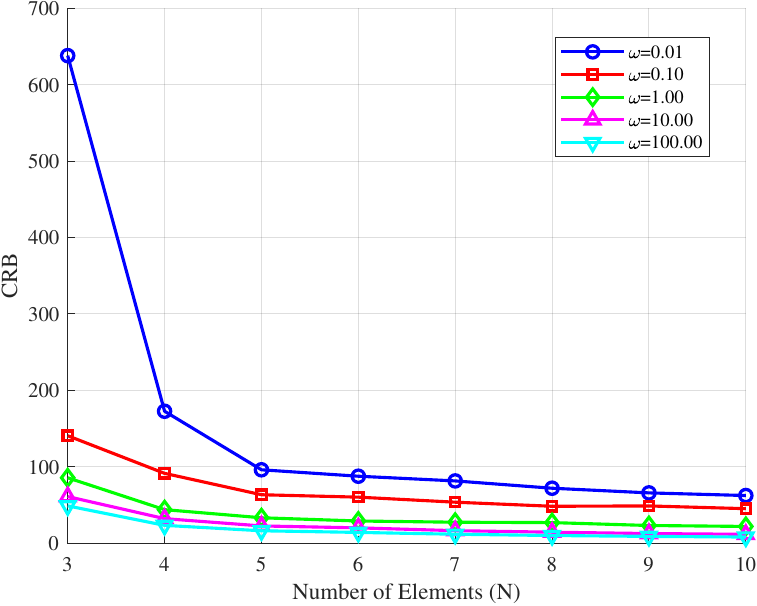}
	\caption{CRB performance versus swarm size $N$ under different values of the weight parameter $\omega$.}
	\label{fig:omega_crb}
\end{figure}

Fig.~\ref{fig:omega_rate} depicts the achievable rate versus the swarm size $N$ for different $\omega$.  
A smaller $\omega$ (i.e., prioritizing communication) yields higher rates, while a larger $\omega$ results in lower rates as the optimization increasingly emphasizes sensing accuracy.  
The corresponding CRB performance is shown in Fig.~\ref{fig:omega_crb}, where a larger $\omega$ significantly reduces the CRB, confirming its role in enhancing localization accuracy at the cost of throughput.  

\begin{figure}[t]
	\centering
	\includegraphics[width=0.85\columnwidth]{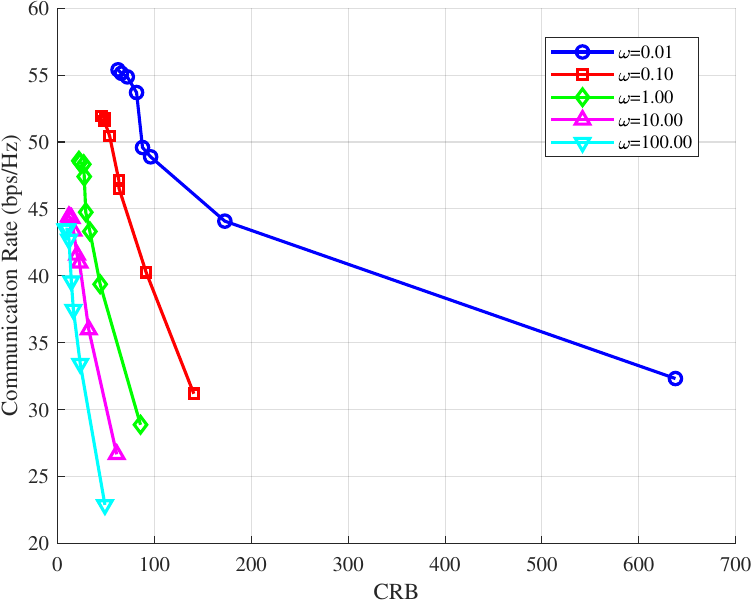}
	\caption{Pareto trade-off curves between rate and CRB under different values of the weight parameter $\omega$.}
	\label{fig:pareto}
\end{figure}

Finally, Fig.~\ref{fig:pareto} presents the Pareto trade-off curves between the communication rate and CRB for different values of $\omega$.  
Each curve corresponds to a distinct weighting configuration, collectively forming the achievable rate–sensing frontier.  
These results demonstrate that no single configuration simultaneously optimizes both metrics; rather, the weight parameter $\omega$ offers a flexible means to tune the system behavior according to specific ISAC task requirements.  

\subsection{Comparison with State-of-the-Art Baselines}
{
To further validate the effectiveness of the proposed UAV swarm-enabled ISAC optimization framework, we first note that existing UAV swarm studies employing virtual antenna arrays  mainly focus on enhancing communication performance, such as capacity or coverage optimization~\cite{hanna2021uav,9771817,10195219}, without considering the joint sensing functionality. 
Therefore, to the best of our knowledge, there are no existing UAV swarm-based ISAC schemes available for direct comparison. 
For performance evaluation, we compare our proposed framework with two representative baseline schemes:}
\begin{itemize}
	\item \textbf{Uniform distribution:} UAVs are evenly located within a $100\times100\times100~\mathrm{m}^3$ cubic region without optimization.
	\item \textbf{Random distribution:} UAVs are randomly deployed in the same cubic region, without any position adjustment.
\end{itemize}

\begin{figure}[t]
	\centering
	\includegraphics[width=0.85\columnwidth]{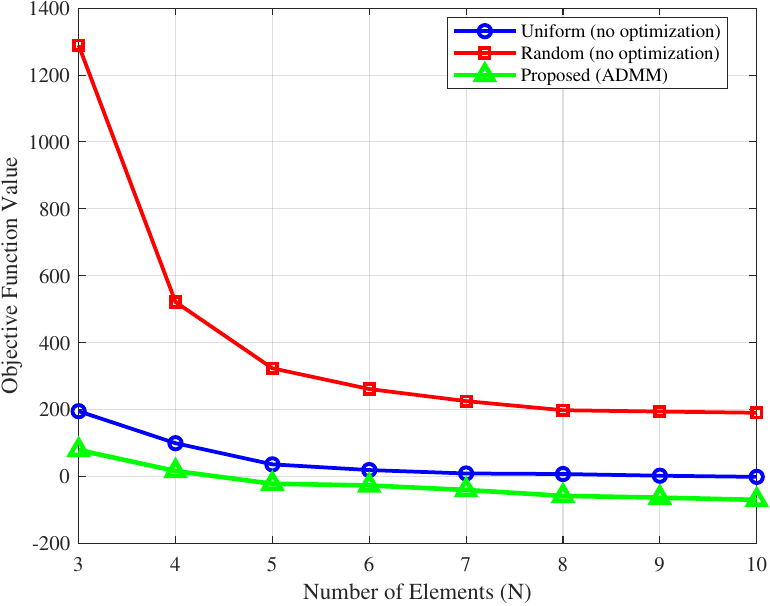}
	\caption{Comparison of the objective function under three schemes: uniform distribution, random distribution, and the proposed consensus ADMM.}
	\label{fig:comparison_objective}
\end{figure}

Fig.~\ref{fig:comparison_objective} presents the objective function value in \eqref{eq:ISAC_opt} for different swarm sizes $N$ under the three schemes.  
The proposed consensus ADMM-based approach consistently achieves the lowest objective value, demonstrating its superior ability to balance communication throughput and sensing accuracy.  
Compared with the random baseline, the improvement is significant; a clear gain is observed compared with the uniform baseline.

Unlike many array optimization scenarios where the performance gap grows with system scale, the absolute improvement of the proposed method remains relatively stable as $N$ increases.  
This is because enlarging the swarm benefits all schemes by introducing additional spatial diversity, thereby improving both communication and sensing performance even without optimization.  
Nevertheless, our approach maintains a consistent advantage across all tested swarm sizes, highlighting its robustness and scalability in large-scale UAV swarms.

\section{Conclusions}{
This paper studied a UAV swarm-enabled ISAC system and proposed a decentralized algorithm that allows UAVs to determine their positions and reach consensus on a globally optimal swarm geometry. 
Two performance metrics, i.e., achievable uplink rate and CRB, were derived to measure communication and sensing, respectively. 
The new decentralized consensus ADMM algorithm was designed to combine local gradient updates with limited inter-UAV information exchange. 
Through iterative consensus steps, the UAVs align their local estimates and reach consensus.
Closed-form gradients for both metrics were derived for efficient implementation and fast convergence. 
Extensive simulations verified that the decentralized consensus ADMM algorithm achieves a superior rate–sensing trade-off compared to baselines, and exhibits robust scalability across different swarm sizes.
}

	\appendices

\section{Proof of Proposition 1}\label{appa}

The received signal at the UAV swarm is modeled as
\begin{equation*}
	\mathbf{y} = \mathbf{H} \mathbf{x} + \mathbf{n},
\end{equation*}
where $\mathbf{y} \in \mathbb{C}^{N}$ is the received signal vector, $\mathbf{H} \in \mathbb{C}^{N \times M}$ denotes the uplink channel matrix between the user and the UAV swarm, $\mathbf{x} \in \mathbb{C}^{M}$ is the transmitted signal vector, and $\mathbf{n} \sim \mathcal{CN}(\mathbf{0}, \sigma^2 \mathbf{I}_N)$ represents the additive white Gaussian noise.

Assuming that the transmitted signal $\mathbf{x}$ follows a complex Gaussian distribution with zero mean and covariance $\mathbf{U} = \mathbb{E}[\mathbf{x} \mathbf{x}^H]$, the capacity of this uplink MIMO channel is given by the mutual information expression \cite{tse2005fundamentals}:
\begin{equation*}
	R = \log \det \left( \mathbf{I}_N + \frac{1}{\sigma^2} \mathbf{H} \mathbf{U} \mathbf{H}^H \right).
\end{equation*}
When the transmitter has no CSI and adopts a spatially white input, i.e., $\mathbf{U} = \mathbf{I}_M$, the capacity simplifies to
\begin{equation*}
	R = \log \det \left( \mathbf{I}_N + \frac{1}{\sigma^2} \mathbf{H} \mathbf{H}^H \right),
\end{equation*}
which corresponds to the achievable uplink rate of the UAV swarm-enabled ISAC system.

\section{Proof of Proposition 2}

We estimate the user’s global position $\mathbf{p}_u \in \mathbb{R}^3$ based on the propagation delay measurements collected by the $N$ UAVs in the swarm from the $M$ transmit antennas of the user.  
The propagation delay between the $m$-th transmit antenna (located at $\mathbf{p}_u + \boldsymbol{\delta}_m$) and the $n$-th UAV (located at $\mathbf{q}_n$) is given by
\begin{equation*}
	\tau_{n,m}(\mathbf{p}_u) = \frac{\left\| \mathbf{q}_n - (\mathbf{p}_u + \boldsymbol{\delta}_m) \right\|}{c}.
\end{equation*}
Due to measurement noise, the observed delay is modeled as
\begin{equation*}
	z_{n,m} = \tau_{n,m}(\mathbf{p}_u) + w_{n,m},
\end{equation*}
where $w_{n,m} \sim \mathcal{N}(0, \varsigma^2)$ is i.i.d. Gaussian noise.  
Let $\mathbf{z} \in \mathbb{R}^{NM}$ be the stacked observation vector of all delay measurements.

The FIM with respect to the unknown user position $\mathbf{p}_u$ is defined as
\begin{equation*}
	\mathbf{J}(\mathbf{p}_u) = \mathbb{E}_{\mathbf{z}|\mathbf{p}_u} \!\left[ 
	\left( \frac{\partial \log p(\mathbf{z}; \mathbf{p}_u)}{\partial \mathbf{p}_u} \right)
	\left( \frac{\partial \log p(\mathbf{z}; \mathbf{p}_u)}{\partial \mathbf{p}_u} \right)^\top 
	\right],
\end{equation*}
where $p(\mathbf{z}; \mathbf{p}_u)$ denotes the likelihood of observing $\mathbf{z}$ given the user position.

Since each observation $z_{n,m}$ follows a Gaussian distribution with mean $\tau_{n,m}(\mathbf{p}_u)$ and variance $\varsigma^2$, its conditional PDF is
\begin{equation*}
	p(z_{n,m}; \mathbf{p}_u) = \frac{1}{\sqrt{2\pi \varsigma^2}} 
	\exp\!\left( -\frac{1}{2\varsigma^2} (z_{n,m} - \tau_{n,m}(\mathbf{p}_u))^2 \right).
\end{equation*}
Assuming independent measurement noises $\{w_{n,m}\}$, the log-likelihood of all observations becomes
\begin{align*}
	\log p(\mathbf{z}; \mathbf{p}_u)
	= & -\frac{1}{2\varsigma^2} \sum_{n=1}^N \sum_{m=1}^M 
	\left( z_{n,m} - \tau_{n,m}(\mathbf{p}_u) \right)^2 \nonumber \\ 
	& - \frac{NM}{2} \log(2\pi \varsigma^2).
\end{align*}
Taking the gradient of $\log p(\mathbf{z}; \mathbf{p}_u)$ with respect to $\mathbf{p}_u$ yields
\begin{align*}
	\frac{\partial \log p(\mathbf{z}; \mathbf{p}_u)}{\partial \mathbf{p}_u} 
	= \frac{1}{\varsigma^2} \sum_{n=1}^N \sum_{m=1}^M 
	\left( z_{n,m} - \tau_{n,m} \right)
	\cdot \left( - \nabla_{\mathbf{p}_u} \tau_{n,m} \right).
\end{align*}
Taking the expectation of the outer product of this score function gives
\begin{align*}
	\mathbf{J}(\mathbf{p}_u)
	&= \frac{1}{\varsigma^2} 
	\sum_{n=1}^N \sum_{m=1}^M
	\mathbb{E} \!\left[ (z_{n,m} - \tau_{n,m})^2 \right] \nonumber \\
	&\quad \times 
	\left( \nabla_{\mathbf{p}_u} \tau_{n,m} \right)
	\left( \nabla_{\mathbf{p}_u} \tau_{n,m} \right)^\top.
\end{align*}
Since $\mathbb{E}[z_{n,m}] = \tau_{n,m}$ and $\mathbb{E}[(z_{n,m} - \tau_{n,m})^2] = \varsigma^2$, the FIM simplifies to
\begin{align*}
	\mathbf{J}(\mathbf{p}_u) = \frac{1}{\varsigma^2} \sum_{n=1}^N \sum_{m=1}^M 
	\left( \nabla_{\mathbf{p}_u} \tau_{n,m}(\mathbf{p}_u) \right)
	\left( \nabla_{\mathbf{p}_u} \tau_{n,m}(\mathbf{p}_u) \right)^\top.
\end{align*}
By the chain rule, we have
$\nabla_{\mathbf{p}_u} \tau_{n,m}(\mathbf{p}_u) 
= - \frac{\mathbf{d}_{n,m}}{c \|\mathbf{d}_{n,m}\|}$, where 
$  \mathbf{d}_{n,m} = \mathbf{q}_n - (\mathbf{p}_u + \boldsymbol{\delta}_m)$.
Defining the unit vector $\mathbf{u}_{n,m} = \frac{\mathbf{d}_{n,m}}{\|\mathbf{d}_{n,m}\|}$, we can express the FIM as
\begin{equation}
	\mathbf{J}(\mathbf{p}_u) = \frac{1}{\varsigma^2 c^2} \sum_{n=1}^{N} \sum_{m=1}^{M} \mathbf{u}_{n,m} \mathbf{u}_{n,m}^\top. \label{aa10}
\end{equation}
The corresponding CRB, which provides a lower bound on the mean squared error (MSE) of any unbiased position estimator, is thus given by
\begin{equation*}
	\mathrm{CRB}(\mathbf{p}_u) = \operatorname{Tr}\!\left( \mathbf{J}^{-1}(\mathbf{p}_u) \right),
\end{equation*}
which, substituting \eqref{aa10}, can be written as
\begin{equation*}
	\mathrm{CRB}(\mathbf{p}_u) = \operatorname{Tr} \!\left( \left( \frac{1}{\varsigma^2 c^2} 
	\sum_{n=1}^{N} \sum_{m=1}^{M} \mathbf{u}_{n,m} \mathbf{u}_{n,m}^\top \right)^{-1} \right).
\end{equation*}

\section{Proof of Proposition 3}\label{app_c}
\subsection{Gradient of Regularization Term}

In the consensus update step of the UAV swarm-enabled ISAC optimization, the quadratic regularization term is 
\begin{equation*}
	g(\{ \mathbf{z}_n \}) 
	= \frac{\rho}{2} \sum_{n=1}^N 
	\left\| \mathbf{q}_n^{k+1} - \mathbf{z}_n + \frac{1}{\rho}\boldsymbol{\mu}_n^k \right\|^2.
	\label{eq:quad_term}
\end{equation*}
For a given UAV $n$ in the swarm, the corresponding local sub-function is expressed as
\begin{equation*}
	g_n(\mathbf{z}_n) 
	= \frac{\rho}{2} \left\| \mathbf{q}_n^{k+1} - \mathbf{z}_n + \frac{1}{\rho}\boldsymbol{\mu}_n^k \right\|^2.
\end{equation*}
Expanding the squared norm yields
\begin{align*}
	g_n(\mathbf{z}_n) 
	&= \frac{\rho}{2} 
	\big(\mathbf{q}_n^{k+1} - \mathbf{z}_n + \tfrac{1}{\rho}\boldsymbol{\mu}_n^k \big)^\top
	\big(\mathbf{q}_n^{k+1} - \mathbf{z}_n + \tfrac{1}{\rho}\boldsymbol{\mu}_n^k \big) \notag \\
	&= \frac{\rho}{2} \left( 
	\mathbf{z}_n^\top \mathbf{z}_n 
	\!-\! 2 \mathbf{z}_n^\top \Big(\mathbf{q}_n^{k+1} \!\!+\! \tfrac{1}{\rho}\boldsymbol{\mu}_n^k \Big)
	\!+ \!\big\| \mathbf{q}_n^{k+1} \!\!+\! \tfrac{1}{\rho}\boldsymbol{\mu}_n^k \big\|^2
	\right).
\end{align*}
Taking the gradient of $g_n(\mathbf{z}_n)$ with respect to $\mathbf{z}_n$, we obtain
\begin{align*}
	\nabla_{\mathbf{z}_n} g_n(\mathbf{z}_n) 
	&= \frac{\rho}{2} \left( 2\mathbf{z}_n - 2\Big(\mathbf{q}_n^{k+1} + \tfrac{1}{\rho}\boldsymbol{\mu}_n^k \Big) \right) \notag \\
	&= \rho \mathbf{z}_n - \rho \mathbf{q}_n^{k+1} - \boldsymbol{\mu}_n^k.
\end{align*}

\subsection{Gradient of Communication Rate}\label{app_d}

We start from the  achievable communication rate:
\begin{equation*}
	R(\mathbf{Z}) = \log \det \!\left( \mathbf{I}_N + \frac{1}{\sigma^2}\mathbf{H}(\mathbf{Z}) \mathbf{H}^H(\mathbf{Z}) \right),
\end{equation*}
where $\mathbf{H}(\mathbf{Z}) \in \mathbb{C}^{N\times M}$ denotes the uplink channel matrix from the user to the UAV swarm, which depends on the UAV positions $\mathbf{Z}=\{\mathbf{z}_n\}$.  

Let
\begin{equation}
	\mathbf{A} = \mathbf{I}_N + \frac{1}{\sigma^2}\mathbf{H}\mathbf{H}^H.\label{60}
\end{equation}
Then, the rate can be rewritten as
\begin{equation}
	R(\mathbf{Z}) = \log \det(\mathbf{A}).\label{61}
\end{equation}
A standard result from matrix calculus states that, for any invertible matrix $\mathbf{A}$,
\begin{equation}
	d \log \det (\mathbf{A}) = \mathrm{Tr}\!\left(\mathbf{A}^{-1} d\mathbf{A}\right).\label{62}
\end{equation}
Substituting \eqref{62} into \eqref{61} yields
	$dR = \mathrm{Tr}\!\left(\mathbf{A}^{-1} d\mathbf{A}\right)$.
On the other hand, \eqref{60} leads to
	$d\mathbf{A} = \frac{1}{\sigma^2}\left(d\mathbf{H}\mathbf{H}^H + \mathbf{H}d\mathbf{H}^H\right)$,
and, in turn, 
	$dR = \frac{1}{\sigma^2}\mathrm{Tr}\!\left(\mathbf{A}^{-1} d\mathbf{H}\mathbf{H}^H\right) 
	+ \frac{1}{\sigma^2}\mathrm{Tr}\!\left(\mathbf{A}^{-1} \mathbf{H} d\mathbf{H}^H\right)$.

Applying the cyclic property of the trace, $\mathrm{Tr}(\mathbf{X}\mathbf{Y}) = \mathrm{Tr}(\mathbf{Y}\mathbf{X})$, we can rewrite
\begin{align*}
	\mathrm{Tr}\!\left(\mathbf{A}^{-1} d\mathbf{H}\mathbf{H}^H\right) 
	&= \mathrm{Tr}\!\left(\mathbf{H}^H \mathbf{A}^{-1} d\mathbf{H}\right), \\
	\mathrm{Tr}\!\left(\mathbf{A}^{-1}\mathbf{H} d\mathbf{H}^H\right) 
	&= \mathrm{Tr}\!\left(d\mathbf{H}^H \mathbf{A}^{-1} \mathbf{H}\right).
\end{align*}
Define
	$\mathbf{X} = \mathbf{H}^H \mathbf{A}^{-1} d\mathbf{H}$.
Then,
$\mathbf{X}^H = d\mathbf{H}^H \mathbf{A}^{-1} \mathbf{H}$.
Since $\mathrm{Tr}(\mathbf{X}^H) = [{\mathrm{Tr}(\mathbf{X})}]^*$, the two terms combine as
\begin{align}
	dR &= \frac{1}{\sigma^2}\left( \mathrm{Tr}(\mathbf{X}) + [{\mathrm{Tr}(\mathbf{X})}]^* \right) \notag\\
	&= \frac{2}{\sigma^2}\Re\!\left\{ \mathrm{Tr}(\mathbf{H}^H \mathbf{A}^{-1} d\mathbf{H}) \right\}.\label{32}
\end{align}
By the definition of complex matrix gradients, the differential of $R$ can also be expressed as
\begin{equation}
	dR = \Re\!\left\{ \mathrm{Tr}\!\left( (\nabla_{\!\mathbf{H}} R)^H d\mathbf{H} \right) \right\}.
	\label{66}
\end{equation}
Comparing \eqref{32} and \eqref{66}, we obtain
	$\nabla_{\!\mathbf{H}} R = \frac{2}{\sigma^2}\mathbf{A}^{-1}\mathbf{H}$.

Next, we analyze the gradient of the communication rate with respect to the UAV position $\mathbf{z}_n$.  
By expanding the trace in \eqref{66}, it readily follows that 
\begin{equation*}
	\mathrm{Tr}\!\left( (\nabla_{\!\mathbf{H}} R)^H d\mathbf{H} \right) 
	= \sum_{i=1}^N \sum_{m=1}^M \left[(\nabla_{\!\mathbf{H}} R)\right]_{i,m}^* \, d[\mathbf{H}]_{i,m}.
\end{equation*}
Thus, \eqref{32} can be rewritten as
\begin{equation}
	dR = \Re\!\left\{ \sum_{i=1}^N \sum_{m=1}^M \big[\nabla_{\!\mathbf{H}} R\big]_{i,m}^* \, d[\mathbf{H}]_{i,m} \right\}.\label{37}
\end{equation}
The variation of each channel coefficient depends on the UAV positions and can be expressed as
	$d[\mathbf{H}]_{i,m} = \frac{\partial [\mathbf{H}]_{i,m}(\mathbf{Z})}{\partial \mathbf{z}_n} \, d\mathbf{z}_n$,
which is nonzero only when $i=n$.  
Hence, \eqref{37} becomes
\begin{equation}
	dR = \Re\!\left\{ \sum_{i=1}^N \sum_{m=1}^M 
	\big[\nabla_{\!\mathbf{H}} R\big]_{i,m}^* 
	\frac{\partial [\mathbf{H}]_{i,m}(\mathbf{Z})}{\partial \mathbf{z}_n} \, d\mathbf{z}_n \right\}.\label{38}
\end{equation}
Since $\tfrac{\partial [\mathbf{H}]_{i,m}}{\partial \mathbf{z}_n}=0$ for all $i\neq n$, \eqref{38} simplifies to
\begin{equation*}
	dR = \Re\!\left\{ \sum_{m=1}^M 
	\big[\nabla_{\!\mathbf{H}} R\big]_{n,m}^* 
	\frac{\partial [\mathbf{H}]_{n,m}(\mathbf{Z})}{\partial \mathbf{z}_n} \, d\mathbf{z}_n \right\}.
\end{equation*}
Comparing this with the definition
	$dR = \Re\!\left\{ (\nabla_{\mathbf{z}_n} R)^T d\mathbf{z}_n \right\}$,
we finally obtain
\begin{equation}
	\nabla_{\mathbf{z}_n} R(\mathbf{Z}) = 
	\sum_{m=1}^M \Re\!\left\{ \big[\nabla_{\!\mathbf{H}} R\big]_{n,m}^* 
	\frac{\partial [\mathbf{H}]_{n,m}(\mathbf{Z})}{\partial \mathbf{z}_n} \right\}.\label{aa1}
\end{equation}

We proceed to compute $\frac{\partial [\mathbf{H}]_{n,m}}{\partial \mathbf{z}_n}$.  
Let $g(r) = \beta_0 r^{-\gamma} \exp\!\left(-j\frac{2\pi}{\lambda}r\right)$.
Then, $[\mathbf{H}]_{n,m} = g(r_{n,m})$. By the chain rule,  it follows that
	$\frac{\partial [\mathbf{H}]_{n,m}}{\partial \mathbf{z}_n} = \frac{dg}{dr}\Big|_{r=r_{n,m}} \cdot \frac{\partial r_{n,m}}{\partial \mathbf{z}_n}$,
where
	$\frac{\partial r_{n,m}}{\partial \mathbf{z}_n}
	= \frac{\partial r_{n,m}}{\partial r_{n,m}^2} \cdot \frac{\partial r_{n,m}^2}{\partial \mathbf{z}_n} 
    = \frac{\mathbf{d}_{n,m}}{r_{n,m}}$,
and
\begin{align*}
	\frac{dg}{dr} &= \beta_0 \left( \frac{d}{dr}\left( r^{-\gamma} \right)\cdot e^{-j\frac{2\pi}{\lambda}r} 
	+ r^{-\gamma} \cdot \frac{d}{dr}\left(e^{-j\frac{2\pi}{\lambda}r}\right)\right) \\
	&= \beta_0 \left(-\frac{\gamma}{r^{\gamma+1}} - j\frac{2\pi}{\lambda}\frac{1}{r^\gamma}\right) e^{-j\frac{2\pi}{\lambda}r}.
\end{align*}
As a result, we have
\begin{equation}
	\frac{\partial [\mathbf{H}]_{n,m}}{\partial \mathbf{z}_n} 
	\!\!= \!\!\beta_0 \Big(-\!\frac{\gamma}{r_{n,m}^{\gamma+2}}\! - \!j\frac{2\pi}{\lambda}\frac{1}{r_{n,m}^{\gamma+1}}\Big)
	e^{-j\frac{2\pi}{\lambda}r_{n,m}}{\mathbf{d}_{n,m}}.\label{39}
\end{equation}

By substituting \eqref{39} into \eqref{aa1}, the gradient of the communication rate with respect to $\mathbf{z}_n$ is
given in \eqref{eq:grad_rate_final}.

\subsection{Gradient of CRB}\label{app_e}

We derive the gradient of the CRB with respect to the UAV positions, $\mathbf{z}_n \in \mathbb{R}^3$. 
The CRB is expressed as
\begin{equation*}
	\mathrm{CRB}(\mathbf{Z}) = \varsigma^2 c^2 \,\mathrm{Tr}(\mathbf{J}(\mathbf{Z})^{-1}).
\end{equation*}
Using the differential identity $d(\mathbf{J}^{-1}) = -\mathbf{J}^{-1}(d\mathbf{J})\mathbf{J}^{-1}$, it readily follows that
\begin{equation}
	d\,\mathrm{CRB}(\mathbf{Z}) 
	= \varsigma^2 c^2 \,d\,\mathrm{Tr}(\mathbf{J}^{-1}) 
	= -\varsigma^2 c^2 \,\mathrm{Tr}\!\left(\mathbf{J}^{-1}(d\mathbf{J})\mathbf{J}^{-1}\right).
	\label{eq:CRB_diff}
\end{equation}
Since $\mathbf{J}(\mathbf{Z})$ depends on all UAV positions $\{\mathbf{z}_1,\dots,\mathbf{z}_N\}$, its differential can be written as
\begin{equation}
	d\mathbf{J} = \sum_{n=1}^N \nabla_{\mathbf{z}_n} \mathbf{J}(\mathbf{Z})\, d\mathbf{z}_n,
	\label{aa3}
\end{equation}
Substituting \eqref{aa3} into \eqref{eq:CRB_diff}, we obtain
\begin{align*}
	d\,\mathrm{CRB}(\mathbf{Z}) 
	&= -\varsigma^2 c^2 \,\mathrm{Tr}\!\left(\mathbf{J}^{-1}\left(\sum_{n=1}^N \nabla_{\mathbf{z}_n} \mathbf{J}(\mathbf{Z})\,d\mathbf{z}_n\right)\mathbf{J}^{-1}\right) \\
	&= \sum_{n=1}^N \left(-\varsigma^2 c^2 \,\mathrm{Tr}\!\left(\mathbf{J}^{-1}(\nabla_{\mathbf{z}_n} \mathbf{J}(\mathbf{Z})\,d\mathbf{z}_n)\mathbf{J}^{-1}\right)\right).
\end{align*}
For any UAV $n$ in the swarm, the corresponding term is
\begin{equation*}
	dF_n = -\varsigma^2 c^2 \,\mathrm{Tr}\!\big(\mathbf{J}^{-1}(\nabla_{\mathbf{z}_n} \mathbf{J}(\mathbf{Z})\,d\mathbf{z}_n)\mathbf{J}^{-1}\big).
\end{equation*}

The third-order tensor $\nabla_{\mathbf{z}_n} \mathbf{J}(\mathbf{Z})$ is defined elementwise as
\begin{equation*}
	\big[\nabla_{\mathbf{z}_n} \mathbf{J}(\mathbf{Z})\big]_{ij,k} = \frac{\partial J_{ij}(\mathbf{Z})}{\partial (\mathbf{z}_n)_k}, 
	\quad i,j,k=1,2,3.
\end{equation*}
Thus, for UAV $n$, we have
\begin{equation}
	(\nabla_{\mathbf{z}_n} \mathbf{J}(\mathbf{Z})\, d\mathbf{z}_n)_{ij} = \sum_{k=1}^3 \frac{\partial J_{ij}(\mathbf{Z})}{\partial (\mathbf{z}_n)_k} (d\mathbf{z}_n)_k,
	\label{aa2}
\end{equation}
which is a $3\times 3$ matrix increment caused by the perturbation of $\mathbf{z}_n$.
Hence, we have
\begin{equation}
	dF_n = \sum_{k=1}^3 \left(-\varsigma^2 c^2 \,\mathrm{Tr}\!\Big(\mathbf{J}^{-1}\frac{\partial \mathbf{J}}{\partial (\mathbf{z}_n)_k}\mathbf{J}^{-1}\Big)\right) (d\mathbf{z}_n)_k.
	\label{aa9}
\end{equation}

On the other hand, by the definition of the gradient, we have
\begin{equation}
	dF_n = \sum_{k=1}^3 \frac{\partial \,\mathrm{CRB}(\mathbf{Z})}{\partial (\mathbf{z}_n)_k} (d\mathbf{z}_n)_k.
	\label{aa5}
\end{equation}
Comparing \eqref{aa9} and \eqref{aa5}, we can readily conclude that
\begin{equation*}
	\frac{\partial \,\mathrm{CRB}(\mathbf{Z})}{\partial (\mathbf{z}_n)_k} 
	= -\varsigma^2 c^2 \,\mathrm{Tr}\!\Big(\mathbf{J}^{-1}\frac{\partial \mathbf{J}}{\partial (\mathbf{z}_n)_k}\mathbf{J}^{-1}\Big),
	\quad k=1,2,3.
\end{equation*}
Thus, the gradient of the CRB with respect to $\mathbf{z}_n$ is given by
\begin{equation}
	\nabla_{\mathbf{z}_n}\,\mathrm{CRB}(\mathbf{Z}) =
	\begin{bmatrix}
		-\varsigma^2 c^2 \,\mathrm{Tr}\!\big(\mathbf{J}^{-1}\tfrac{\partial \mathbf{J}}{\partial (\mathbf{z}_n)_1}\mathbf{J}^{-1}\big) \\
		-\varsigma^2 c^2 \,\mathrm{Tr}\!\big(\mathbf{J}^{-1}\tfrac{\partial \mathbf{J}}{\partial (\mathbf{z}_n)_2}\mathbf{J}^{-1}\big) \\
		-\varsigma^2 c^2 \,\mathrm{Tr}\!\big(\mathbf{J}^{-1}\tfrac{\partial \mathbf{J}}{\partial (\mathbf{z}_n)_3}\mathbf{J}^{-1}\big)
	\end{bmatrix}.
	\label{eq:CRB_grad_general}
\end{equation}

Now, it remains to expand the partial derivative $\tfrac{\partial \mathbf{J}}{\partial (\mathbf{z}_n)_k}$. Since
$
	\mathbf{J}(\mathbf{Z}) = \sum_{i=1}^N \sum_{m=1}^M \mathbf{u}_{i,m}\mathbf{u}_{i,m}^\top$,
only the terms with $i=n$ depend on $\mathbf{z}_n$. Thus,
\begin{equation}
	\frac{\partial \mathbf{J}}{\partial (\mathbf{z}_n)_k} 
	= \sum_{m=1}^M \frac{\partial (\mathbf{u}_{n,m}\mathbf{u}_{n,m}^\top)}{\partial (\mathbf{z}_n)_k}.
	\label{aa13}
\end{equation}

Next, we compute the derivative of $\mathbf{u}_{n,m}$. Let
\begin{equation*}
	\mathbf{d}_{n,m} = \mathbf{z}_n - (\mathbf{p}_u+\boldsymbol{\delta}_m), 
	\quad r_{n,m} = \|\mathbf{d}_{n,m}\|,
	\quad \mathbf{u}_{n,m} = \frac{\mathbf{d}_{n,m}}{r_{n,m}}.
\end{equation*}
The derivative of $(\mathbf{u}_{n,m})_i$ with respect to $(\mathbf{z}_n)_k$ is given by
\begin{align*}
	\frac{\partial (\mathbf{u}_{n,m})_i}{\partial (\mathbf{z}_n)_k}
	&= \frac{\delta_{ik} r_{n,m} - (\mathbf{d}_{n,m})_i \tfrac{(\mathbf{d}_{n,m})_k}{r_{n,m}}}{r_{n,m}^2} \\
	&= \frac{1}{r_{n,m}} \left(\delta_{ik} - (\mathbf{u}_{n,m})_i (\mathbf{u}_{n,m})_k\right),
\end{align*}
where $\delta_{ik}$ is the Kronecker delta. Noting that $(\mathbf{e}_k)_i=\delta_{ik}$, the result can be compactly written as
\begin{equation}
	\frac{\partial \mathbf{u}_{n,m}}{\partial (\mathbf{z}_n)_k} 
	= \frac{1}{r_{n,m}} \Big(\mathbf{I} - \mathbf{u}_{n,m}\mathbf{u}_{n,m}^\top\Big)\mathbf{e}_k,
	\label{aa11}
\end{equation}
where $\mathbf{e}_k \in \mathbb{R}^3$ denotes the $k$-th standard basis vector.

Applying the product rule to the outer product $\mathbf{u}_{n,m}\mathbf{u}_{n,m}^\top$, it follows that
\begin{equation}
	\frac{\partial (\mathbf{u}_{n,m}\mathbf{u}_{n,m}^\top)}{\partial (\mathbf{z}_n)_k}
	= \Big(\frac{\partial \mathbf{u}_{n,m}}{\partial (\mathbf{z}_n)_k}\Big)\mathbf{u}_{n,m}^\top 
	+ \mathbf{u}_{n,m}\Big(\frac{\partial \mathbf{u}_{n,m}}{\partial (\mathbf{z}_n)_k}\Big)^\top.
	\label{aa123}
\end{equation}
Substituting \eqref{aa11} into \eqref{aa123} gives
\begin{align}
	\frac{\partial (\mathbf{u}_{n,m}\mathbf{u}_{n,m}^\top)}{\partial (\mathbf{z}_n)_k}
	&= \frac{1}{r_{n,m}}
	\Big[(\mathbf{I}-\mathbf{u}_{n,m}\mathbf{u}_{n,m}^\top)\mathbf{e}_k \mathbf{u}_{n,m}^\top \nonumber \\
	&\quad + \mathbf{u}_{n,m} \mathbf{e}_k^\top (\mathbf{I}-\mathbf{u}_{n,m}\mathbf{u}_{n,m}^\top)\Big].
	\label{aa14}
\end{align}

Finally, substituting \eqref{aa13} and \eqref{aa14} into \eqref{eq:CRB_grad_general}, we obtain the gradient of the CRB with respect to $\mathbf{z}_n$, as given in \eqref{28}.

\bibliographystyle{IEEEtran}
	\bibliography{IEEEabrv,mybib_ieee_fixed}
\end{document}